\documentclass[aps,pre,preprint,groupedaddress,
,superscriptaddress,showpacs,a4paper,floatfix]{revtex4-1}

\usepackage{amsmath,amssymb,graphicx,graphics,bm}
\usepackage{hyperref}
\usepackage{graphics,graphicx}
\usepackage{amsmath,amssymb}

\begin{document}

\title{Fluid-particle interactions and fluctuation-dissipation relations II -
Gaussianity and Gaussianity breaking}

\author{Chiara Pezzotti}
\email{chiara.pezzotti@unroma1.it}
\affiliation{Facolt\`{a} di Ingegneria, La Sapienza Universit\`{a} di Roma,
via Eudossiana 18, 00184, Roma, Italy}

\author{Massimiliano Giona}
\email{massimiliano.giona@uniroma1.it}
\affiliation{Facolt\`{a} di Ingegneria, La Sapienza Universit\`{a} di Roma,
via Eudossiana 18, 00184, Roma, Italy}

\author{Giuseppe Procopio}
\email{giuseppe.procopio@uniroma1.it}
\affiliation{Facolt\`{a} di Ingegneria, La Sapienza Universit\`{a} di Roma,
via Eudossiana 18, 00184, Roma, Italy}

\date{\today}

\begin{abstract}
The analysis of fluctuation-dissipation relations developed in  
Giona et al. (2024)  for particle hydromechanics is
extended to stochastic forcings alternative to Wiener processes, 
with the aim of addressing the occurrence
of Gaussian equilibrium densities or alternatively the breaking of the Gaussian
paradigm at equilibrium.
Preliminarly, it is discussed how the determination of the
fluctuational patterns starting from the Gaussian approach to Markov processes
is practically unfeasible, and the moment analysis provides the
simplest way to  achieve it. 
We show  the existence of an uncountable family of white-noise processes, 
different from the distributional derivatives of Wiener processes,
and satisfying the  requirement of fluctuational independence, i.e.
the basic assumption on thermal fluctuations  in the Kubo  theory
based on the Langevin condition.
The importance of this extension is that it may provide a transition from
mesoscopic to microscopic (event-based)  stochastic modeling.
In this framework,   the derivatives  of Wiener processes
constitute a very peculiar, albeit continuous, element of this class.
The fluctuational patterns driven
by non-Wiener stochastic forcings   
display in general non-Gaussian velocity fluctuations at equilibrium,
and the Gaussian case is recovered in the limit of small perturbations.
Finally, a fully hydromechanic approach to anomalous diffusion is provided,
both in the subdiffusive and in the superdiffusive cases.

\end{abstract}

\maketitle

\section{Introduction}
\label{2sec1}
In  \cite{part1}, henceforth referred to as part I, we have developed a simple and complete formulation of the
Fluctuation-Dissipation  (FD) relations, focused on 
the explicit and analytic representation 
of particle hydromechanics in any linear complex fluid and flow conditions,
  consistent with the basic assumption
of  the fluctuational independence  of the stochastic perturbations \cite{kubo,kubolibro}. 
This has led to the introduction of the concept of
fluctuational patterns driven by distributional
derivatives of Wiener processes. Particularly important in this approach is the boundedness
of fluid-inertial memory kernels, deriving from viscoelastic modeling
of the fluid and stemming from the finite propagation velocity of
shear stresses \cite{procopvisco}.

The stochastic representation of the thermal force is
 referred to as the  Fluctuation-Dissipation relation of the third kind (FD3k, for short).
Within the
decomposition of the hydrodynamic effects into
 dissipative and fluid-inertial contributions,
 the most ``conservative'' assumption regarding the elementary
stochastic processes  in the  representation of the FD3k
 has been assumed in part I, by choosing  them
 as distributional derivatives of Wiener processes (Wiener ansatz).

Albeit this assumption is commonly accepted, as it is in line with the classical stochastic
formulation of statistical physics \cite{ystatphys1,ystatphys2} and its
extension to non-equilibrium phenomena \cite{onsager1,onsager2}, 
it still remains a  mathematical ansatz
to be justified and argumented, case by case, on physical grounds.
The Wiener ansatz  leads necessarily to  Gaussian equilibrium 
probability density functions (pdf)
 \cite{strato}. Nevertheless,
other, physically plausible models are possible, and this article  addresses this issue. This is particularly important in order to connect the classical
thermodynamics  and transport theory developed for fluid phases
in ``non-extreme conditions'' (ambient temperature, atmospheric pressure, etc.)
with the models of athermal systems that display anomalies with respect to the
Gaussian paradigm \cite{athermal1,athermal2},
underlying the analogies and pinpointing the differences.

Consider the motion of a spherical particle in a quiescent fluid 
(Brownian motion) \cite{langevin,chandra,vankampen}.  In
linear hydrodynamic regimes (corresponding either to
the Stokes regime, where the response of the fluid is instantaneous,
or  to the time-dependent Stokes regime in which the inertia 
of the fluid is accounted for by explicitly considering the derivative
of the  fluid velocity  in the hydrodynamic equations) \cite{landauf}, it is described by the evolution equation for the particle velocity $v(t)$
\begin{equation}
m \, \frac{d v(t)}{d t}= {\mathcal L}[v(t)] + R(t)
\label{2eq1_1}
\end{equation}
where ${\mathcal L}[v(t)]$ is a linear functional
of the velocity and $R(t)$ the stochastic  thermal force, as extensively
addressed in part I. The basic
statistical condition in the assessment of fluctuation-dissipation relations
is that the thermal force $R(t)$ is statistically independent of $v(t^\prime)$ for any
$t^\prime\leq t$ \cite{kubo,kubolibro}.
This leads  to the Langevin condition 
\begin{equation}
\langle R(t) \, v(t^\prime) \rangle =0 \, , \qquad t^\prime \leq t
\label{2eq1_2}
\end{equation}
that is also the main assumption in Linear Response Theory (LRT).
From eq. (\ref{2eq1_2}), the evolution of the velocity autocorrelation
function  follows, and  it  is possible to derive out of it the
conditions for the modal stochastic contributions, i.e. for
the elementary stochastic forcings acting in the fluctuational
patterns,  addressed in part I.
The latter conditions  are  automatically fulfilled if the  modal stochastic forces  are
expressed as distributional derivatives of independent Wiener processes. However, this choice
does not constitute the unique representation for them,  and the
analysis and  the implications of this fact  are one of the 
main  goals of the present article.
In point of fact,  the non-uniqueness in the representation of the stochastic
thermal forces opens up new physical issues, essentially suggesting a more
detailed characterization of the statistics of particle fluctuations
\cite{gionaklagesfd}.

The introduction of other classes of stochastic forcing  deviating from
the Wiener ansatz can be easily imbedded within the fluctuational
patterns. The  conceptual importance of this extension is that
it makes it possible to generalize the analysis of fluctuations from a
mesoscopic level, associated with the ``collective description''
encompassed by the Wiener ansatz and stemming from Central Limit Theorem (CLT), to
a microscopic level, where the fluctuations describe elementary random events.
This issue is addressed in Sections \ref{2sec2} and \ref{2sec3}.
The main phenomenological implication of this extension is that the
equilibrium velocity pdfs are no longer Gaussian, and Gaussianity can be recovered in
the limit of small perturbations.

The scope of this article is to analyze the relation between FD theory and
equilibrium  Gaussianity (in the meaning stated above). Obviously,
the extension to other non-Wiener classes of stochastic forcings is a central
issue in this regard. For the sake of completeness two other topics should be also
carefully addressed.
The first one is the methodological use of Gaussianity in deriving FD3k, i.e. the
use of the Gaussian nature of equilibrium densities to enforce from it the values
of the internsities of the stochastic forcings. This is analyzed in Section \ref{2sec2} and it is
shown that the moment approach, thoroughly developed in part I, is, by far, the simplest
and most streamlined way to achieve this goal.

The other crucial point is the extension of particle hydromechanics beyond
the regular-diffusion case, i.e. to address fluid-particle interactions leading to
an emergent anomalous diffusive motion for which the mean square displacement deviates
from a long-term linear scaling with time. Anomalous behaviours have been observed in many complex and crowded media \cite{Franosch, Fedotov1, Regner}. The formulation
of a global fluctuation-dissipation relation
in the anomalous case (subdiffusive motion) is due to Mason
and coworkers \cite{mason1,mason2,mason3} (see also \cite{gser1,gser2,gser3})
and is customarily referred to as the Generalized Stokes-Einstein Relation
(GSER). Recent advancements have further extended the theoretical underpinnings and applications of the GSER. For instance, McKinley and Nguyen \cite{McKinley2018} demonstrated how the structure of the memory kernel in the GLE can determine the long-term subdiffusive behaviour of the mean square displacement, providing a rigorous framework for analyzing viscoelastic diffusion. Ferreira \cite{Ferreira2022} proposed a scaling method for systems with memory, offering new analytical tools to classify and interpret different diffusion regimes, including subdiffusion and superdiffusion, within the GSER context. Furthermore, Tuladhar et al. \cite{Tuladhar2017} investigated anomalous relaxations driven by non-Poisson renewal processes, showing how distinct memory structures influence the dynamics, both from a GLE perspective and through alternative approaches like subordination.

In this article we focus on the hydromechanic theory of anomalous
motion, starting from the properties of the dissipative and fluid-inertial kernels,
$h(t)$ and $k(t)$ introduced in part I, and the resulting equilibrium velocity pdf.
While, as expected, subdiffusive motion is recovered in the presence of long-tailed
dissipative kernels the time-integrals of which diverge, superdiffusive motion
necessarily implies the absence of dissipation and long-tailed fluid-inertial kernels.

The article is organized as follows. Section \ref{2sec2} discusses 
the occurrence of Gaussian distributions in the case of the Wiener-based formulation of the 
FD3k  addressed in part I. It briefly  reviews the classical formulation of 
fluctuation-dissipation relations for Gaussian Markovian processes 
\cite{strato}, and
the  differences with respect to the moment approach developed in part I.
Section \ref{2sec3} analyzes the occurrence of other classes of stochastic
processes  fulfilling the requirement of  fluctuational independence  eq. (\ref{2eq1_2}) \cite{vankampen},
corresponding to  the distributional derivatives of
Markovian or semi-Markovian compound counting processes \cite{semimarkovcount1,semimarkovcount2}.
Section \ref{2sec4} develops the FD3k relations in the case of Markovian processes,
 while Section \ref{2sec5} analyzes
the extensions to semi-Markov processes. 
Several examples are analyzed, and the Gaussianity breaking 
of the associated pdf  discussed in detail.  
Section \ref{2sec6} unifies all the cases  treated in Sections \ref{2sec4} and \ref{2sec5}  
as a generalization of the 
 the fluctuational patterns defined in part I.
Finally, Section \ref{2sec7} develops the hydromechanic theory of anomalous
diffusion.

\section{Gaussianity and FD}
\label{2sec2}

All the fluctuational patterns analyzed in part I are driven by distributional
derivatives of Wiener processes and lead to  Gaussian equilibrium pdfs.
To make an example,  consider the case of  dissipative memory effects in the
nondimensional formulation in which the equilibrium velocity variance is normalized to 1 (see part I for a detailed
description of this model arising from a dissipative memory kernel $h(t)=\sum_{h=1}^{N_d} a_h \, e^{-\lambda_h \, t}$, $a_h, \, \lambda_h >0$, $h=1,\dots,N_d$), the dynamic equations of which are expressed by
\begin{eqnarray}
\frac{d v}{d t} & = & - \sum_{h=1}^{N_d} a_h \, \theta_h \nonumber \\
\frac{d \theta_h}{d t}& = & - \lambda_h \, \theta_h+ v+ \sqrt{2} \, b_h \, \xi_h(t) \, , \quad h=1,\dots,N_d
\label{2eq2_1}
\end{eqnarray}
where the fulfillment of the Langevin condition eq. (\ref{2eq1_2}) provides 
$b_h = \sqrt{\lambda_h/a_h}$, $h=1,\dots,N_d$. In this case, since
the equilibrium mixed moments of the dissipative degrees of freedom $\{\theta_h \}_{h=1}^{N_d}$
are given by
\begin{equation}
m_{\theta_h \theta_k}^*= \sigma_h^2 \, \delta_{h,k} \, , \qquad \sigma_h^2 = \frac{1}{a_h}
\label{2eq2_2}
\end{equation}
and since $m_{vv}^*=1$ at equilibrium, it is easy to prove, from
the properties of the resulting Fokker-Planck equation, that the
equilibrium distribution $p^*(v,{\boldsymbol \theta})$   attains the form
\begin{equation}
p^*(v,{\boldsymbol \theta})= A \, e^{-v^2/2} \prod_{h=1}^{N_d}
e^{-\theta_h^2/2 \, \sigma_h^2}
\label{2eq2_3}
\end{equation}
where $A$ is a normalization constant. In a similar 
way, in the presence of both dissipative and fluid-inertial memory effects,
described by the internal dissipative $\{\theta_h\}_{h=1}^{N_d}$,
and fluid-inertial $\{ z_\alpha \}_{\alpha=1}^{N_i}$  variables, the nondimensional equations of motion
attain the  form
\begin{eqnarray}
\frac{d v}{d t} & = &- \left ( \sum_{\alpha=1} \gamma_\alpha \right ) v - \sum_{h01}^{N_d} a_h \, \theta_h + \sum_{\alpha=1}^{N_i} \gamma \, \mu_\alpha \, z_\alpha + \sqrt{2} \, \sum_{\alpha=1}^{N_i}
d_\alpha \, \xi_\alpha^\prime(t) \nonumber \\
\frac{d \theta_h}{d t} & =  &- \lambda_h \, \theta_h + v + \sqrt{2} \, b_h \, \xi_h(t) 
\label{2eq2_4} \\
\frac{d z_\alpha}{d t} &=  &- \mu_\alpha \, z_\alpha + v + \sqrt{2} \, c_\alpha \, \xi_\alpha^\prime(t)
\nonumber
\end{eqnarray}
The  parameters $b_h$, $c_\alpha$ and $d_\alpha$ modulating the intensity of the Wiener forcings are
derived in part I from eq. (\ref{2eq1_2})   
 and  the equilibrium density  $p^*(v,{\boldsymbol \theta},{\bf z})$ is
also Gaussian
\begin{equation}
p^*(v,{\boldsymbol \theta},{\bf z})= A \, e^{-v^2/2} \, \prod_{h=1}^{N_d}
e^{-\theta_h^2/2 \sigma_h^2} \, \prod_{\alpha=1}^{N_i} e^{-z_\alpha^2/2 \, s_\alpha^2}
\label{2eq2_6}
\end{equation}
where  the $\sigma_h^2$'s are given by eq. (\ref{2eq2_2}) and
\begin{equation}
s_\alpha^2 = \frac{c_\alpha^2}{\mu_\alpha}= \frac{1}{\gamma_\alpha \, \mu_\alpha} \,  \quad
\alpha=1,\dots,N_i
\label{2eq2_7}
\end{equation}
\subsection{FD for Gaussian Markovian processes}
\label{2sec2_1}

In the light of these properties, the analysis developed in  part I 
 for determining the FD3k representation could be in principle deduced from the 
FD relations for Markovian processes analyzed by Stratonovich \cite{strato},
which are briefly reviewed below.

Consider the Fokker-Planck equation for a generic Markovian model,
defined by the state variable ${\bf y}=(y_1,\dots,y_n)$ and
expressed by a second-order parabolic equation for the
pdf $p({\bf y},t)$,
\begin{equation}
\frac{\partial p}{\partial t}= \sum_{i=1}^n \frac{\partial }{\partial y_i}
\left ( \sum_{j=1}^n d_{ij} \, y_j \, p \right ) +
\frac{1}{2} \sum_{i,j=1}^n K_{ij} \frac{\partial^2 p}{\partial y_i \partial y_j}
\label{2eq2_8}
\end{equation}
where a linear deterministic contribution (in our case stemming from
hydrodynamic interactions) defined by the  matrix ${\bf d}$,
the entries  of which are $d_{ij}$,  has been included.
The matrix ${\bf d}$ is positive semidefinite i.e.,
all the eigenvalues admit non negative real part.
The Fokker-Planck equation (\ref{2eq2_8})
 admits a Gaussian equilibrium distribution $p^*({\bf y})$,
\begin{equation}
p^*({\bf y}) = A \, \exp \left [ -\frac{1}{2} \sum_{i,j=1}^n a_{ij} y_i \, y_j \right ]
\label{2eq2_9}
\end{equation}
where the matrix ${\bf a}$, with entries $a_{ij}$, is related to the  symmetric matrix 
${\bf M}$, $M_{ij}= \langle y_i \, y_j \rangle_{\rm eq}=m_{y_i y_j}^*$ 
of the second-order moments at equilibrium by the relation
\begin{equation}
{\bf a}= {\bf M}^{-1}
\label{2eq2_10}
\end{equation}
Substituting eq. (\ref{2eq2_9}) into eq. (\ref{2eq2_8}) at steady state,
the expression for the ``diffusivity'' matrix ${\bf K}$ (we use, freely this
diction although the entries of  ${\bf K}$ do not possess the physical dimensions of a square-length per time)
 follows
\begin{equation}
K_{ij}= \sum_{h=1}^n d_{ih} \, M_{hj} + \sum_{h=1}^n d_{jh} \, M_{ih}
\label{2eq2_11}
\end{equation}
or in matrix form
\begin{equation}
{\bf K}= {\bf d} \, {\bf M} + {\bf M} \, {\bf d}^T
\label{2eq2_12}
\end{equation}
where ``$T$'' indicates transpose. Eq. (\ref{2eq2_12}) represents the
more general FD relation according to Stratonovich.

\subsection{Application to FD3k}
\label{2sec2_2}

In principle, one could apply this result in order to derive  FD3k, 
i.e. a specific representation of the
stochastic perturbation as addressed by Stratonovich.
The Stratonovich route to FD3k, that can be referred to as the Gaussian approach reads as
follows: given ${\bf d}$, i.e. the hydromechanic $N \times N$ interaction matrix,
assume a functional form of the moment matrix ${\bf M}$ consistent with
the Langevin condition, such that the resulting ``diffusivity matrix ${\bf K}$ defined
by eq. (\ref{2eq2_12}) would be symmetric and positive semidefinite. Once a candidate  for
${\bf K}$ fulfilling these two properties has been found, the stochastic
perturbation defining FD3k follows simply as ${\bf K}^{1/2} \,  \boldsymbol{\xi}(t)$,
where $\boldsymbol{\xi}(t)=(\xi_1,\dots,\xi_N(t))$ is
the distributional derivative of a $N$-dimensional Wiener process.

 However, the implementation of this approach is not
so simple, as problems arise in the
practical application of this formalism in order to derive the structure of the stochastic perturbation.
To make a simple example, consider the case of dissipative memory
in the presence of a single mode, where
\begin{equation}
{\bf y}= 
\left (
\begin{array}{c}
y_1 \\
y_2
\end{array}
\right ) =
\left (
\begin{array}{c}
v \\
\theta_1
\end{array}
\right ) \, , \qquad
{\bf d}= 
\left (
\begin{array}{cc}
0  & 1 \\
-1 \; &  \; \lambda_1 
\end{array}
\right )
\label{2eq2_13}
\end{equation}
and $\lambda_1 >0$.
Since $\langle y_1^2\rangle_{\rm eq}
 = \langle v^2 \rangle_{\rm eq}=1$, and $\langle y_1 \, y_2 \rangle_{\rm eq}= 
\langle v \, \theta_1 \rangle_{\rm eq}=0$, the moment matrix ${\bf M}$
is given by
\begin{equation}
{\bf M} =
\left (
\begin{array}{cc}
1 \;& 0 \\
0 \; &  \; \sigma^2
\end{array}
\right )
\label{2eq2_14}
\end{equation}
where $\sigma^2=\langle y_2^2\rangle_{\rm eq} = \langle \theta_1^2 \rangle_{\rm eq}$
is the squared variance of the dissipative degree of freedom $\theta_1$
at equilibrium, and is a parameter to be determined. In principle $\sigma^2$ could attain any positive value,
and given $\sigma^2$, the expression for the diffusivity
matrix ${\bf K}$ follows. In the present case, from eqs. (\ref{2eq2_12})-(\ref{2eq2_13}) the matrix ${\bf K}$ attains the form 
\begin{equation}
{\bf K}= \left  (
\begin{array}{cc}
\; 0 \; & \; \sigma^2-1 \; \\
\sigma^2-1 &  \; 2 \, \lambda_1 \, \sigma^2 \;
\end{array}
\right )
\label{2eq2_15}
\end{equation}
Both from physical and mathematical reasons the matrix ${\bf K}$
should be positive semidefinite as otherwise: (i) an equilibrium solution
could not
exist; (ii) it would not exist a  square $2 \times 2$
matrix ${\bf b}$ such that ${\bf K}={\bf b}^2$, and consequently
it would not exist a stochastic representation of the
dynamics
\begin{equation}
d {\bf y} = - {\bf d} \, dt + {\bf  b } \, d {\boldsymbol \xi}(t)
\label{2eq2_16}
\end{equation}
where ${\boldsymbol \xi}(t)=(\xi_1(t),\xi_2(t))$,  the entries
of which are distributional derivatives of independent Wiener
processes. Indeed, eq. (\ref{2eq2_16}), with the specification 
of the matrix ${\bf b}$, represents
in this case the solution of the FD3k-problem.
Since $\mbox{Tr}({\bf K})=2  \, \lambda_1  \, \sigma^2>0$,
and $\mbox{Det}({\bf K})=- (\sigma^2 -1)^2$, the matrix ${\bf K}$ is not
positive semidefinite unless that for $\sigma=1$, for which it reduces
to 
\begin{equation}
{\bf K}= \left  (
\begin{array}{cc}
\; 0 \; &  \; 0 \; \\
\; 0 \; & 2 \, \lambda_1 
\end{array}
\right ) \qquad \Rightarrow
\qquad
{\bf b}= \left  (
\begin{array}{cc}
\; 0 \; &  \; 0 \; \\
\; 0 \; &  \sqrt{2 \, \lambda_1}
\end{array}
\right )
\label{2eq2_17}
\end{equation}
and this provides the correct representation of the dynamics
as derived in part I.

The generalization of this simple case  to a spectrum of relaxation
rates and to the inclusion of fluid inertial effects 
shows clearly the  practical
shortcomings of this approach.
In a  higher-dimensional
case, the moments are not known {\em a-priori},
and their  admissible values should
be determined by the condition that the diffusivity matrix ${\bf K}$ should be positive semidefinite in order to  satisfy  the physical condition of
existence of an equilibrium  state and to
provide a stochastic  model for the process.
This approach, either in the dissipative case when $N_d$ becomes large
and ${\bf y}=(v,{\boldsymbol \theta})$ or {\em a fortiori}
in the dissipative/inertial case
where  both $N_d$ and $N_i$ are large and 
${\bf y}=(v,{\boldsymbol \theta},{\bf z})$ becomes extremely difficult, 
if not unfeasible, to pursue.
Moreover, it cannot be applied in the presence of a continuous spectrum
of relaxation rates.

On the other hand, the direct representation of the stochastic
dynamics in terms of evolution equations
for $v$, ${\boldsymbol \theta}$, ${\bf z}$ ($v$, ${\boldsymbol \theta}$,
in the purely dissipative case) see part I for details, 
provides a simple, general and safe way for
approaching and solving the FD3k problem for  stochastic particle hydromechanics. 

\section{White noise: Wiener and impulsive processes}
\label{2sec3}

The formal definition of Langevin equations  in the presence of thermal fluctuations has been addressed by van Kampen \cite{vankampen} restricting the nature of the stochastic process $R(t)$.
Van Kampen conditions, 
apart from the assumption of stochasticity for $R(t)$ (from \cite{vankampen} 
p.219 ``..{\em This $L(t)$} (cfr. corresponding to $R(t)$ in eq. (\ref{2eq1_1}))
{\em is irregular and unpredictable.....Thus $L(t)$ can be treated as a stochastic
process.}''),  involve the condition
\begin{equation}
\langle R(t) \rangle =0
\label{2eq4_0}
\end{equation}
i.e. a unbiased action of the
thermal fluctuations due to homogeneity and isotropy,
and
\begin{equation}
\langle R(t) \, R(t^\prime) \rangle = R_0 \, \delta(t-t^\prime)
\label{2eq4_1}
\end{equation}
i.e. the $\delta$-correlated  character of $R(t)$, where $R_0$ is a constant.
From  the principle of causality and  from the linearity
of eq. (\ref{2eq1_1}), $v(t^\prime)$  is a linear
functional of the history of noise from $t=0$ up to $t=t^\prime$,
and consequently if eq. (\ref{2eq4_1}) is satisfied, it follows
that eq. (\ref{2eq1_2}) is also fulfilled (with some
caution in the case $t=t^\prime$, that can be treated invoking
non-anticipativity \cite{gardiner}).
Eqs. (\ref{2eq4_0})-(\ref{2eq4_1})  imply that $R(t)$ should belong to the class
of   white noise processes \cite{gardiner,white2}.

A subsequent analysis on the fundamental structural properties
of the random force in  the Langevin equations, has been  developed
by  
Mazur  and Bedeaux \cite{bedeaux1,bedeaux2}. 
Considering the scalar Langevin equation
\begin{equation}
m \, \frac{d v}{d t}= B(v) + R(t)
\label{eq24_2}
\end{equation}
where $B(v)$ could be in general nonlinear, with the
 conditions that there exists an equilibrium distribution $P_0(v)$, and
$\langle B(y) \rangle = \int B(v) \, P_0(v) \, d v =0$, these
authors claim that enforcing causality and time-reversal invariance,
``{\em ...a Langevin force (cfr. $R(t)$) which is independent of the state of the system is necessarily  white} (cfr. i.e. it  belongs
 to the class of white noises)
{\em and Gaussian.}''  While the white-noise condition, coinciding with the
analysis  developed by van Kampen  eqs. (\ref{2eq4_0})-(\ref{2eq4_1}),   is correct, the claim
for Gaussianity invoked by the authors is   overstated, 
for the simple reason that 
 it does not
depend on the above mentioned assumptions, but rather on
a further ansatz put forward by Mazur and Bedeaux \cite{bedeaux1,bedeaux2}, 
namely that
the equilibrium pdf $P_0(y)$ should be related to the deterministic
part of the force, i.e. to $B(v)$, by the condition 
\begin{equation}
\frac{d \, \log P_0(v)}{d v} \propto B(v)
\label{2eq4_3}
\end{equation}
where "$\propto$" indicates strict proportionality.
If the Mazur-Bedeaux result would be of general validity  under the
weaker assumptions mentioned above, this would
imply  that the only admissible $R(t)$ are proportional to the  distributional
derivatives of  a Wiener process, which is Gaussian and white.
Hovewer, the  claim of Gaussianity  in the  argument by  Mazur and Bedeaux
is epistemically circular, resambling a  
 tautology. To make an example, 
 for $B(v)=-\eta \, v$, i.e. for the Stokesian
friction, if the equilibrium distribution should be Gaussian (corresponding to 
the Maxwellian
velocity distribution), the noise should be also Gaussian, which is nothing but
the closure condition of  Gaussian  Markovian stochastic processes 
in the presence of linear forcings addressed in section
\ref{2sec2}. But eq. (\ref{2eq4_3}) is not a principle of nature, and indeed
it is inconsistent with respect to fundamental physical theories, such
as relativity, for which the  Maxwellian  distribution is the  low-energy
approximation of a class of more fundamental distributions, i.e.
the J\"uttner distribution \cite{hanggirel}.  Consequently,
 Gaussianity
cannot be invoked solely for causality and time-reversal invariance.

For this
reason, we follow exclusively the van Kampen conditions.
In point of fact, as addressed above, the van Kampen conditions correspond
exactly to the principle of stochastic independence invoked in the
LRT, indicating that $R(t)$, and as well all the
modal stochastic forcings $\xi_h(t)$, $\xi_\alpha^\prime(t)$ 
entering the fluctuational patterns developed in part I
 should
be white-noise processes.

The distributional derivative of a Wiener process is just one particular member
of this class, and other, equal legitimate alternatives
are possible,  the FD-analysis of which is addressed in the remainder.
Specifically, one can consider for the elementary stochastic forcings
$\xi_h(t)$, $\xi_\alpha^\prime(t)$ the distributional derivatives of a normalized and unbiased compound Poisson process $\chi(t,\nu;\beta)$  \cite{compoundpoisson} where $\nu$ is the constant transition rate and $\beta$  a random variable corresponding to the
increment of the process at each transition event, defined
by its pdf $f_\beta(\beta)$. The conditions of unbias and normalization
imply for the first order moments of $\beta$
\begin{equation}
\langle \beta \rangle = \int_{-\infty}^\infty \beta \, f_\beta(\beta) \, d \beta=0 \, , \qquad 
\langle \beta^2 \rangle =
\int_{-\infty}^\infty \beta^2 \, f_\beta(\beta) \, d \beta =1
\label{2eq4_5}
\end{equation}
Let $\eta(t,\nu;\beta)=d \chi(t,\nu;\beta)/d t$. In this case we have
\begin{equation}
\langle \eta(t,\nu; \beta) \, \eta(t^\prime,\nu; \beta)=  \nu \, \delta(t-t^\prime)
\label{2eq4_6}
\end{equation}
The process $\eta(t,\nu;\beta)$ can be conceptually represented
as
\begin{equation}
\eta(t,\nu;\beta)= \sum_{h=1}^\infty \beta_h \, \delta(t-t_h)
\label{2eq4_7}
\end{equation}
with $t_h< t_{h+1}$,
where $\tau_1=t_1$, and $ \tau_h=t_{h}-t_{h-1}$, $h=2,3,\dots$ are independent and identically
distributed  (iid) random variables, possessing an exponential
density  function $T_{\tau}(\tau)= \nu \, e^{-\nu \, \tau}$, $\tau \in [0,\infty)$, and
the $\beta_h$'s are independent  and identically distributed random
variables defined by the density $f_\beta(\beta)$, $\beta \in {\mathbb R}$,
satisfying eq. (\ref{2eq4_5}).
Due to  the Markovian nature of the process eq. (\ref{2eq4_7})  in the presence
of an exponential distribution $T_\tau(\tau)$, the  statistical
characterization  of its cumulative process $y(t)=\chi(t,\nu;\beta)$
 involves the pdf $P_y(y,t)$ that  is
the solution of the equation
\begin{equation}
\frac{\partial P_y(y,t)}{\partial t}= - \nu \, P_y(y,t) + \int_{-\infty}^\infty f_\beta(\beta^\prime) \, P_y(y-\beta^\prime,t) \, d \beta^\prime
\label{2eq4_8}
\end{equation}
with the initial condition $P_y(y,0)=\delta(y)$. 
In the case the statistics for the  $\tau_h$'s is different from
an exponential distribution, the process defined by eq. (\ref{2eq4_7})
is no longer Markovian and, in order to embed it within
a Markovian representation, an additional variable, namely the transitional
age  $\tau \in [0,\infty)$,  should
be introduced \cite{fedotov,jphysA,extendedPK}. More precisely,
for a generic $T_\tau(\tau)$, define the  age-dependent transition rate $\nu(\tau)$ as
\begin{equation}
T_\tau(\tau)= \nu(\tau) \, \exp \left [ - \int_0^\tau \nu(\theta) \, d \theta)
\right ]
\label{2eq4_9}
\end{equation}
or equivalently
\begin{equation}
\nu(\tau) = \frac{T_\tau(\tau)}{1- \int_0^\tau T_\tau(\theta) \, d \theta}
\label{2eq4_10}
\end{equation}
Let us indicate with $\eta(t,\tau,\nu(\tau);\beta)$ this process and with
$\chi(t,\tau,\nu(\tau);\beta)=\int_0^t \eta(t^\prime,\tau,\nu(\tau);\beta) \, d t^\prime$ its cumulative representation.
The cumulative process $y(t)= \chi(t,\tau,\nu(\tau),\beta)$
is
characterized by the pdf $P(y,\tau,t)$, depending on $\tau$, solution
of the evolution equation
\begin{equation}
\frac{\partial P(y,\tau,t)}{\partial t}= - \frac{\partial P(y,\tau,t)}{\partial \tau} - \nu(\tau) \, P(y,\tau,t)
\label{2eq4_11}
\end{equation}
equipped with the boundary condition
\begin{equation}
P(y,0,t) = \int_{-\infty}^\infty f_\beta(\beta^\prime) \, d \beta^\prime  \int_0^\infty \nu(\tau) \, P(y-\beta^\prime,\tau,t) \, d \tau
\label{2eq4_12}
\end{equation}
and the initial condition $P(y,\tau,0)=\pi_\tau(\tau) \, \delta(y)$,
with $\pi_\tau(\tau) \geq 0$, $\int_0^\infty \pi_\tau (\tau) \, d \tau=1$. If $\nu(\tau)$ defined by eq. (\ref{2eq4_10}) is constant,
the process is Markovian, and $P_y(y,t)= \int_0^\infty P(y,\tau,t) \, d \tau$ is solution of eq. (\ref{2eq4_8}). For generic choices of $T_\tau(\tau)$
different from an exponential distribution, 
the process is semi-Markov. In the case the pdf $T_\tau(\tau)$ possesses
a power law scaling,
\begin{equation}
T_\tau(\tau) = \frac{\xi \, \tau_0^\xi}{(\tau_0+\tau)^{\xi+1}}
\label{2eq4_13}
\end{equation}
where  $\tau_0$ is a characteristic time and $\xi>0$, corresponding to the
transition rate
\begin{equation}
\nu(\tau)= \frac{\xi}{\tau_0 + \tau}
\label{2eq4_14}
\end{equation}
the transitional structure of eq. (\ref{2eq4_7}) is 
analogous to that defining
 L\'evy walks
\cite{levy1}, and giving rise for $\xi\leq 2$ to anomalous superdiffusive
transport. Without loss of generality set $\tau_0=1$.
In the next Sections we analyze the statistical properties of these models,
by discussing first the Markovian case, and subsequently the
semi-Markov extension. Just because of eq. (\ref{2eq4_7}), the class
of Langevin equation (\ref{2eq1_1}) driven by this class of  processes
 can
be referred to as impulse-driven Langevin models.

\section{FD3k for  Markovian impulse-driven processes}
\label{2sec4}

Any Langevin model  eq. (\ref{2eq1_1})  driven by stochastic processes 
satisfying 
eqs. (\ref{2eq4_0})-(\ref{2eq4_1})), and such that at equilibrium  all the internal
degrees of freedom are uncorrelated from the velocity $v$, is a valid 
candidate to express the FD3k relations, and to provide a 
consistent model of microscopic dynamics satisfying Kubo FD relations of the first and second kind \cite{part1,kubolibro}.
In this section we consider the Markovian models  driven by stochastic
forcings defined by eq. (\ref{2eq4_7}), where $\nu$ is a constant,
in the general case of  dissipative/fluid-inertial memory effects, 
considering
the occurrence of a single dissipative and inertial mode, with
internal stochastic perturbation expressed by the impulsive processes
$\eta(t,\nu;\beta)$  introduced in the previous section.
In this case, adopting the block-interaction structure proposed in  
part I, the equations of motion read
\begin{eqnarray}
\frac{d v}{d t} & =& - \gamma \, v - a \, \theta + \gamma \, \mu \, z + q \, \eta_1(t,\nu_1;\beta_1) \nonumber \\
\frac{d \theta}{d t} &=& - \lambda \, \theta + v + b \, \eta_2(t,\nu_2;\beta_2)
\label{2eq5_1} \\
\frac{d z}{d t} & =  &- \mu \, z + v + c \, \eta_1(t,\nu_1;\beta_1) \nonumber
\end{eqnarray}
where $\eta_1$ and $\eta_2$ are two impulsive processes (distributional
derivatives of compound Poisson processes) independent of each other,
and characterized by the constant transition rates $\nu_1$ and $\nu_2$,
respectively, while $\beta_h$, $h=1,2$ are statistically defined by the
pdf's $f_{\beta_h}(\beta_h)$, $h=1,2$ satisfying the normalization
conditions eq. (\ref{2eq4_5}). In eq. (\ref{2eq5_1}) $a$, $\lambda$, $\gamma$ and $\mu$ are given (and positive),
while the constant coefficients $q$, $b$, and $c$ should be determined.

Let $p(v,\theta,z,t)$
be the pdf  for the stochastic dynamics eq. (\ref{2eq5_1}).
Henceforth,  in order to make the notation  more readable, we use the following
convention: given that $(v,\theta,z,t)$ represents
the actual state of the system,
 the bare $p$, without  specifying its
arguments, indicates the pdf evaluated at the actual set of values,
i.e., $p=p(v,\theta,z,t)$, while
$p(v- q \, \beta_1^\prime)=p(v-q \, \beta_1^\prime,\theta,z,t)$  indicates
 that a shift of $-q \, \beta_1^\prime$   is considered in its dependence on the $v$-variable.
In a similar way,
\[
 p(v- q \, \beta_1^\prime, z-c \, \beta_1^\prime)=
p(v-q \, \beta_1^\prime, \theta,z- c\, \beta_1^\prime,t)
\]
Adopting this notation, the Fokker-Planck equation associated with
eq. (\ref{2eq5_1}) reads
\begin{eqnarray}
\frac{\partial p}{\partial t} & = & \gamma \, \frac{\partial \left ( v \, p \right )}{\partial v} +  a \, \theta \, \frac{\partial p}{\partial v} - \gamma \, \mu \, z \, \frac{\partial p}{\partial v} + \lambda \, \frac{\partial \left ( \theta \, p \right )}{\partial \theta}  - v \, \frac{\partial p}{\partial \theta}
 +   \mu \, \frac{ \partial \left ( z \, p \right)}{\partial z} - v \, \frac{\partial p}{\partial z}
 -  \left ( \nu_1 + \nu_2 \right ) \, p  \nonumber \\
&  + & \nu_1 \int_{\mathbb R} f_{\beta_1}(\beta_1^\prime) \, p(v- \, q  \, \beta_1^\prime, z-c \, \beta_1^\prime) \, d \beta_1^\prime + \nu_2 \int_{\mathbb R}  f_{\beta_2}(\beta_2^\prime) \, p(\theta- b \, \beta_2^\prime) \, d \beta_2^\prime
\label{2eq5_2}
\end{eqnarray}
Indicate with $m_v(t)= \langle v(t) \rangle$ the first-order moment of the velocity variables 
(and similarly for $\theta$ and $z$). Since at equilibrium $\langle v \rangle_{\rm eq} = 
\langle \theta \rangle_{\rm eq} = \langle z \rangle_{\rm eq} =0$, the second order moments
provide directly the squared variances and covariances, and
following the approach developed in part I, the FP3k  relations is 
resolved by
enforcing the relations
\begin{equation}
m_{vv}^*= \langle v^2 \rangle_{\rm eq} =1 \,, \quad m_{\theta\,v}^*=\langle \theta v \rangle_{\rm eq} =0 \, , \quad 
m_{z \, v}^*= \langle z v \rangle_{\rm eq} =0
\label{2eq5_3}
\end{equation}
Set $d {\bf v}= d v d\theta d z$ to indicate the measure element with respect
to the state variables.
Making use of the properties eq. (\ref{2eq4_5}), it follows that
\begin{eqnarray}
\int_{{\mathbb R}^3}  d {\bf v} \int_{\mathbb R}  f_\beta(\beta_2^\prime) \, p(\theta- b \, \beta_2^\prime) \, d \beta_2^\prime & = & 1 \nonumber \\
\int_{{\mathbb R}^3}   \theta d {\bf v} \int_{\mathbb R}  f_\beta(\beta_2^\prime) \, p(\theta- b \, \beta_2^\prime) \, d \beta_2^\prime & = & \langle 
\theta + b \, \beta_2 \rangle =0 \label{2eq5_4}\\
\int_{{\mathbb R}^3}   \theta^2 d {\bf v} \int_{\mathbb R}  f_\beta(\beta_2^\prime) \, p(\theta- b \, \beta_2^\prime) \, d \beta_2^\prime & = & 
\langle ( \theta+ b \, \beta_2)^2 \rangle =m_{\theta \theta} + b^2
\nonumber
\end{eqnarray}
and analogously  for the  other integrals,
so that the  equations for the second-order moments  follow. 
Specifically, for $m_{vv}$ we have
\begin{equation}
\frac{d m_{vv}}{d t}= - 2 \, \gamma \, m_{vv}- 2 \, a \, m_{\theta v} + 2 \, \gamma \, \mu \, m_{zv} - (\nu_1+\nu_2) \, m_{vv} + \nu_1 \, \left \langle
(v+ q \, \beta_1)^2 \right \rangle + \nu_2 \, m_{vv}
\label{2eq5_5}
\end{equation}
that at equilibrium, enforcing eqs. (\ref{2eq5_3}), provides
the value of $q$
\begin{equation}
q= \sqrt{\frac{2 \, \gamma}{\nu_1}}
\label{2eq5_6}
\end{equation}
The dynamics of  $m_{\theta \theta}$ is given by
\begin{equation}
\frac{d m_{\theta \theta}}{d t}= 2 \, \lambda \, m_{\theta \theta} + 2 \, 
m_{\theta v} - (\nu_1+\nu_2) \, m_{\theta \theta}  + \nu_1  \, m_{\theta \theta} + \nu_2 \,
\left \langle
(\theta+ b \, \beta_2)^2 \right \rangle 
\label{2eq5_7}
\end{equation}
yielding at equilibrium the value  $m_{\theta \theta}^*$,
\begin{equation}
m_{\theta \theta}^*= \frac{\nu_2 \, b^2}{2 \, \lambda}
\label{2eq5_8}
\end{equation}
while for $m_{z z}$ we have
\begin{equation}
\frac{d m_{z z}}{d t}= - 2 \, \mu \, m_{z z} + 2 \, m_{z v} - (\nu_1 + \nu_2) \, m_{z  z }  + \nu_1  \, \left \langle
(z +  c \, \beta_1)^2 \right \rangle  + \nu_2 \, m_{z z}
\label{2eq5_9}
\end{equation}
that at equilibrium determines the value of $m_{z z}^*$,
\begin{equation}
m_{z z}^*= \frac{ \nu_1 \, c^2}{ 2 \, \mu}
\label{2eq5_10}
\end{equation}
Finally, for the mixed moments we have
\begin{eqnarray}
\frac{d m_{\theta v}}{d t} & = & - \gamma \, m_{\theta v} - a \, m_{\theta \theta} + \gamma \, \mu \, m_{\theta z} - \lambda \, m_{\theta v} + m_{v v}
- (\nu_1 + \nu_2) \, m_{\theta  v}
 \nonumber \\
& + &   \nu_1 \, \left \langle
\theta \, (v+ q \, \beta_1)  \right \rangle + \nu_2   \, \left \langle
 v \, (\theta+ b \, \beta_2) \right \rangle  \nonumber \\
& = & - \gamma \, m_{\theta v} - a \, m_{\theta \theta}
 + \gamma \, \mu \, m_{\theta z}-\lambda \, m_{\theta \, v} + m_{vv}
\label{2ea5_10} \\
\frac{d m_{z v}}{d t} &= &- \gamma \, m_{z v} - a \, m_{\theta z} + \gamma \, \mu \, m_{z z} - \mu \, m_{z v} + m_{v v} - (\nu_1 + \nu_2) \, m_{z  v}  
\nonumber \\
&+  & \nu_1  \, \left \langle
\ (v + q \, \beta_1) \,(z+ c \, \beta_1) \right \rangle + \nu_2  \, 
m_{z v} \nonumber \\
& = & -\gamma \, m_{z v} - a \, m_{\theta z} + \gamma \, \mu \, m_{z z}- \mu \,
m_{z v} + m_{v v} + \nu_1 \, q \, c \label{2eq5_11} \\
\frac{d m_{\theta z}}{d t} & = & - \lambda \, m_{\theta z} + m_{z v} - \mu \, m_{\theta z} +
m_{\theta v} - (\nu_1+ \nu_2 ) \, m_{\theta z}
\nonumber \\
& + &  \nu_1 \, \left \langle
\theta \, (z+ c \, \beta_1) \right \rangle
+  \nu_2 \, \left \langle
z \, (\theta+ b \, \beta_2) \right \rangle \nonumber \\
& = & - \lambda \, m_{\theta z}+ m_{z v} - \mu \, m_{\theta z} + m_{\theta v}
\label{2eq5_12}
\end{eqnarray}
From eq. (\ref{2eq5_12}), enforcing eq. (\ref{2eq5_3}), it follows 
at equilibrium that
\begin{equation}
m_{\theta z}^*=0
\label{2eq5_13}
\end{equation}
and 
\begin{eqnarray}
m_{\theta \theta}^* &=  &\frac{1}{a} \nonumber \\
\gamma  \, \mu \, m_{z z}^*  &+  &1 + \nu_1 \, q \, c  = 0
\label{2eq5_14}
\end{eqnarray}
that, by taking into   account eqs. (\ref{2eq5_6}) and (\ref{2eq5_10}), yields
\begin{equation}
b= \sqrt{\frac{2 \, \lambda}{a \, \nu_2}} \, ,  \qquad
c= - \sqrt{\frac{2}{\gamma \, \nu_1}}
\label{2eq5_15}
\end{equation}
Owing to the uncorrelated relation between the internal degrees of freedom,
this results can be straigthforwardly  generalized 
 to a spectrum of relaxational dissipative and fluid-internial modes,
\begin{eqnarray}
\frac{d v}{d t} & =& -  \sum_{\alpha=1}^{N_i} \gamma_\alpha \, v -  
\sum_{h=1}^{N_d} a _h \, \theta_h +  \sum_{\alpha=1}^{N_i}\gamma_\alpha \, \mu_\alpha \, z_\alpha +   \sum_{\alpha=1}^{N_i}  q_\alpha \eta_\alpha^\prime(t,\nu_\alpha^\prime;\beta_\alpha^\prime) \nonumber \\
\frac{d \theta_h}{d t} &=& - \lambda_h \, \theta_h + v + b_h \, \eta_h(t,\nu_h;\beta_h) \, , \qquad h=1,\dots,N_d
\label{2eq5_16} \\
\frac{d z_\alpha}{d t} & =  &- \mu_\alpha \, z_\alpha + v + c_\alpha \, \eta_\alpha^\prime(t,\nu_\alpha^\prime;\beta_\alpha^\prime)  \, , \qquad \alpha=1,\dots N_i \nonumber
\end{eqnarray}
In the general case, the amplitudes of the stochastic modal  forcings attain
the values
\begin{equation}
q_\alpha= \sqrt{\frac{2 \, \gamma_\alpha}{\nu_\alpha^\prime}} \,, \quad
c_\alpha= - \sqrt{\frac{2}{\gamma_\alpha \, \nu_\alpha^\prime}} \,, \quad \alpha=1,\dots,N_i \,,\quad b_h= \sqrt{\frac{2 \, \lambda_h}{a_h \, \nu_h}} \, , \quad h=1,\dots,N_d
\label{2eq5_17}
\end{equation}

\subsection{Examples}
\label{2sec4_1}

In this paragraph we consider some numerical case studies
 for exemplifying the theory and for addressing
the issue of Gaussianity.
To begin with, consider  the model eq. (\ref{2eq5_1}) with a single dissipative and fluid inertial mode,
$N_d=N_i=1$.
In this case, there are two stochastic contributions $\eta_k(t,\nu_k;\beta)$,  characterized by the transition
rates $\nu_k$, $k=1,2$. The stochastic perturbations  $\eta_k(t,\nu_k;\beta)$ are completely  defined
once the pdf's $f_{\beta_k}(\beta_k)$  for the amplitudes are specified.
The simplest case is a two-state impulsive distribution that, in order to
fulfill the normalization eq. (\ref{2eq4_5}), should  take the expression
\begin{equation}
f_{\beta_k}(\beta_k) = \frac{1}{2} \left [ \delta(\beta_k-1) + \delta(\beta_k+1) \right ] \,,
\quad k=1,2
\label{2eqex1_1} 
\end{equation}
The parameters $q$, $b$ and $c$ entering eq. (\ref{2eq5_1}) follow from the consistency with respect to
the Langevin condition,
i.e.  they are given by eq. (\ref{2eq5_6}) and (\ref{2eq5_15}).

\begin{figure}
\includegraphics[width=10cm]{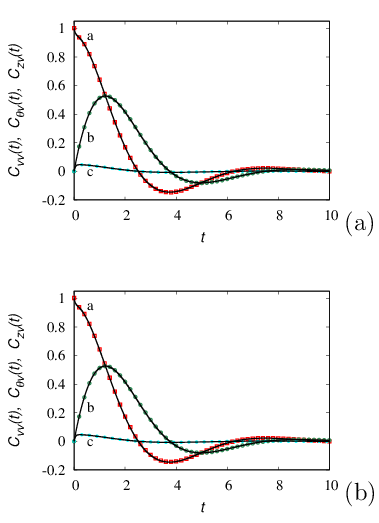}
\caption{Correlation functions  at equilibrium 
for the model eq. (\ref{2eq5_1}), with $a=1$, $\lambda=1$, $\gamma=1$, $\mu=20$ and impulsive
pdf's $f_{\beta_k}(\beta_k)$, $k=1,2$  eq. (\ref{2eqex1_1}).
Symbols represent the results of stochastic simulations, solid lines correspond to the
result of LRT. Panel (a): $\nu_1=0.3$, $\nu_2=0.2$, panel (b): $\nu_1=\nu_2=30$.
Lines (a) refer to $C_{vv}(t)$, lines (b) to $C_{\theta v}(t)$, lines (c) to $C_{zv}(t)$.}
\label{Fig_pk1}
\end{figure}

Figure \ref{Fig_pk1} panels (a) and (b)  show the temporal behaviour of the
correlation functions $C_{vv}(t)$, $C_{\theta v}(t)$ and $C_{z v}(t)$ obtained
from the stochastic simulation of eq. (\ref{2eq5_1}) in the long-time limit (equilibrium conditions),
using an Euler algorithm with  time step $\Delta t$ ranging from $10^{-4}$ to $10^{-3}$, depending on the
value of the transition rates (the higher $\nu_k$'s the smaller $\Delta t$).
The simulations refer to an ensemble of $10^6$ particles, and the averages are performed over $10^9$
realizations. 
The stochastic simulations are compared with the results of LRT for two different
sets of $\nu_k$, $k=1,2$.
Independently of the choice of $\nu_k$, the realizations for fixed properties
of the fluid (i.e. for given dissipative and fluid inertial kernels) provide the same
expression for the correlation functions. This further indicates that, to 
a given particle transport problem, a uncountable family of different FD3k realizations can be associated with.

\begin{figure}
\includegraphics[width=10cm]{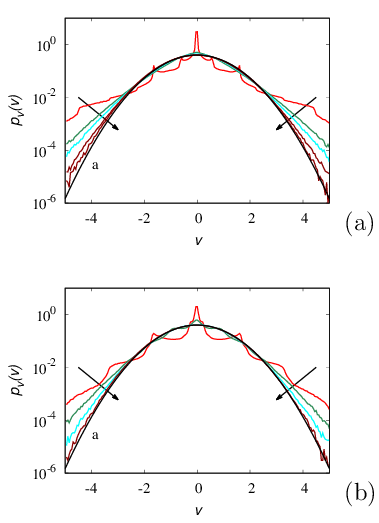}
\caption{Equilibrium velocity pdf $p_v(v)$ vs $v$ for  eq. (\ref{2eq5_1}) using the FD2k relations
eqs. (\ref{2eq5_6}), (\ref{2eq5_15}) in the presence of impulsive
pdf's $f_{\beta_k}(\beta_k)$, $k=1,2$ eq. (\ref{2eqex1_1}). Panel (a) refers to $a=1$, $\lambda=1$, $\gamma=1$, $\mu=1$,
panel (b) to $a=1$, $\lambda=1$, $\gamma=1$, $\mu=20$. The arrows indicate increasing values of $\nu_k$, $k=1,2$.
 Lines (a) represent the normal distribution.
Panel (a): $\nu_2=\nu_1=0.1,\,0.5,\,1,\,3,\,30$, panel (b): $\nu_2=2 \nu_1/3$, $\nu_1=0.3, \, 0.9,\,3,\, 30$.}
\label{Fig_pk2}
\end{figure}

The values of $\nu_1$ and $\nu_2$ determine the shape of the equilibrium
velocity pdf $p_v(v)$, that in the presence of stochastic perturbations different from the derivatives
of Wiener processes, need not to be Gaussian.  The equilibrium pdf $p_v(v)$ are
depicted in figure \ref{Fig_pk2} for two choices of $(a,\lambda,\gamma,\mu)$, by considering
the rates $\nu_1$ and $\nu_2$ as parameters.

The qualitative statistical features and their dependence on $\nu_k$, emerging from
the data  shown in figure \ref{Fig_pk2}  are rather clear.
For small values of $\nu_k$, the equilibrium velocity pdf is characterized by a multispiked shape, and this
is essentially a consequence of the atomic nature of the chosen $f_{\beta_k}(\beta_k)$ eq. (\ref{2eqex1_1}).

As  the $\nu_k$'s increase, the velocity distribution gets smoother and unimodal, attaining
a Gaussian shape in the limit for $\nu_k \rightarrow \infty$, $k=1,2$.
In this limit, the amplitudes $q$, $b$ and $c$ entering eq. (\ref{2eq5_1}) get vanishingly small,
but the products
\begin{equation}
q^2 \, \nu_1 \rightarrow 2 \, \gamma \, , \quad c^2 \, \nu_1 \rightarrow \frac{2}{\gamma} \, ,
\quad b^2 \, \nu_2 \rightarrow \frac{2 \, \lambda}{a}
\label{2eqex1_2}
\end{equation}
attain  constant limit values. The Gaussian limit in this case is 
analogous to the parabolic limit of Poisson-Kac processes, usually referred to as
the Kac limit \cite{kac,gpk1}. Therefore the Gaussian distribution is attained for rapidly varying and
small-amplitude impulsive stochastic forcings.

\begin{figure}
\includegraphics[width=10cm]{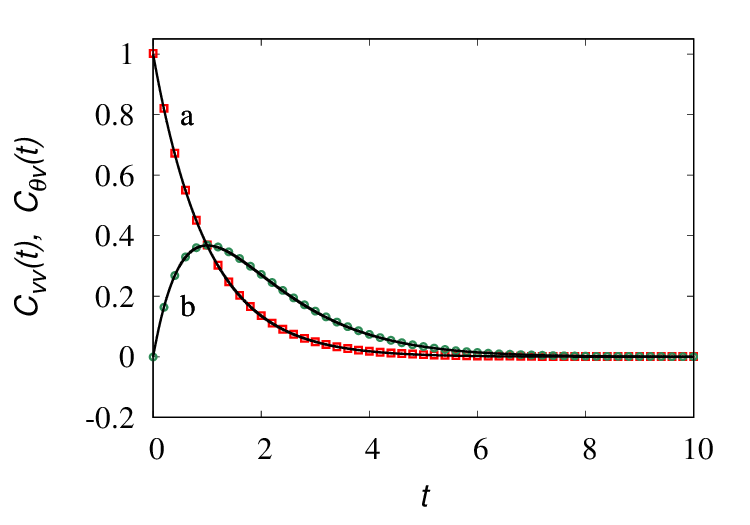}
\caption{Correlation functions  at equilibrium
for the model eq. (\ref{2eq5_1}), with $a=1$, $\lambda=1$, $\gamma=1$, $\mu=1$ and  normal
pdf's $f_{\beta_k}(\beta_k)$, $k=1,2$. Symbols represent the results of stochastic simulations, solid lines correspond to the
result of LRT. Line (a) refers to $C_{vv}(t)$, line (b) to $C_{\theta v}(t)$. 
The correlation function $C_{zv}(t)$ is not depicted as it coincides with $C_{\theta v}(t)$.}
\label{Fig_pk3}
\end{figure}

\begin{figure}
\includegraphics[width=10cm]{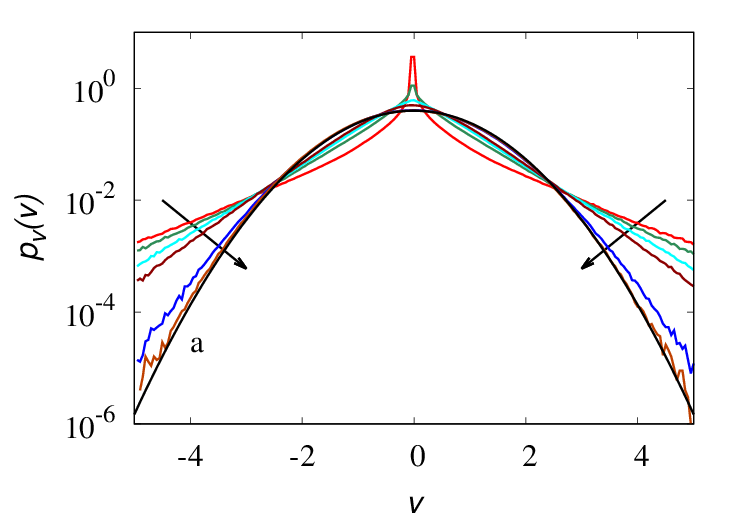}
\caption{Equilibrium velocity pdf $p_v(v)$ vs $v$ for  eq. (\ref{2eq5_1})  with 
$a=1$, $\lambda=1$, $\gamma=1$, $\mu=1$, using the FD2k relations
eqs. (\ref{2eq5_6}), (\ref{2eq5_15}) in the presence of normal 
pdf's $f_{\beta_k}(\beta_k)$, $k=1,2$. The arrows indicate increasing values of $\nu_1=\nu_2=0.1,\,0.3,\,
0.6,\,1,\,10,\,100$. Line (a) represents the normal distribution.} 
\label{Fig_pk4}
\end{figure}

Considering Lebesgue absolutely continuous distributions $f_{\beta_k}(\beta_k)$,
the qualitative physical picture of the role of the parameters defining the elementary
fluctuations, (in the present case the transition rates $\nu_k$, $k=1,2$), remain unaltered.
For instance, consider for $\beta_1$ and $\beta_2$ a normal
distribution, i.e. $f_{\beta_k}(\beta_k)= e^{-\beta_k^2/2}/\sqrt{2 \, \pi}$,
$k=1,2$.
Figures \ref{Fig_pk3} and \ref{Fig_pk4} depict the correlation functions and the
velocity pdf in this case for a given choice of the hydrodynamic parameters $(a,\lambda,\gamma,\mu)$.

The main difference with respect to an atomic distribution of the amplitudes $\beta_k$'s  is
that the resulting equilibrium velocity distributions are unimodal and smoother even for small values of $\nu_k$
compared to the atomic case  depicted in figure \ref{Fig_pk2}, while 
the role of  the $\nu_k$'s governing
the transition towards a Gaussian shape remains unaltered.

These  simple examples clarify  an elementary but important property. For Wiener-driven realizations
of FD3k,  the Gaussianity  of the resulting equilibrium velocity pdf is already built-in
in the statistical structure of the stochastic perturbations (as addressed in sections
\ref{2sec2} and \ref{2sec3}). It is not an emerging thermodynamic property, but rather a 
``statistical  tautology'' owing to the choice of Wiener fluctuations. For other classes
of elementary  modal fluctuations, still satisfying the requirement of
fluctuational independence (see part I for a discussion), Gaussianity, or better to say the Gaussian
approximation, is a resulting  emerging property of the physical interactions for small
amplitudes of the elementary modal stochastic perturbations possessing high transition rates.

\section{FD3k for  semi-Markovian impulse-driven processes}
\label{2sec5}
This section extends the analysis  developed in \ref{2sec4} to the
semi-Markov case.  With reference to eq. (\ref{2eq5_1}), this
implies to replace the stochastic modal forcings $\eta_1(t,\nu_1;\beta_1)$,
$\eta_2(t,\nu_2;\beta_2)$ with their semi-Markov counterparts
$\eta_1(t,\tau_1, \nu_1(\tau_1);\beta_1)$,
$\eta_2(t,\tau_2,\nu_2(\tau_2);\beta_2)$. The resulting pdf $p(v,\theta,z,t;\tau_1,\tau_2)$ depends also on the transitional ages $\tau_1$, $\tau_2$ that should
be introduced in order to describe the semi-Markov transitions, and is the
solution
of the generalized Fokker-Planck equation 
\begin{eqnarray}
\frac{\partial p}{\partial t} & = & \gamma \, \frac{\partial \left ( v \, p \right )}{\partial v} +  a \, \theta \, \frac{\partial p}{\partial v} - \gamma \, \mu \, z \, \frac{\partial p}{\partial v} + \lambda \, \frac{\partial \left ( \theta \, p \right )}{\partial \theta}  - v \, \frac{\partial p}{\partial \theta}
 +   \mu \, \frac{ \partial \left ( z \, p \right)}{\partial z} - v \, \frac{\partial p}{\partial z} \nonumber \\
& - & \frac{\partial  p}{\partial \tau_1} - \frac{\partial p}{\partial \tau_2}
-  \left [ \nu_1(\tau_1) + \nu_2(\tau_2) \right ] \, p  
\label{2eq6_1}
\end{eqnarray}
equipped with the boundary conditions
\begin{equation}
p(v,\theta,z,t;0,\tau_2)= \int_0^\infty \nu_1(\tau_1) \, d \tau_1 \int_{{\mathbb R}} f_1(\beta_1) \,
p(v- q \, \beta_1,\theta,z- c \, \beta_1,t;\tau_1,\tau_2) \, d \beta_1
\label{2eq6_2}
\end{equation}
\begin{equation}
p(v,\theta,z,t;\tau_1,0)= \int_0^\infty \nu_2(\tau_2) \, d \tau_2 \int_{{\mathbb
 R}} f_2(\beta_2) \,
p(v,\theta- b \, \beta_2,z,t;\tau_1,\tau_2) \, d \beta_2
\label{2eq6_3}
\end{equation}
where $f_{\beta_1}(\beta_1)$ and $f_{\beta_2}(\beta_2)$ are such that the random
variables $\beta_1$ and $\beta_2$ satisfy the normalization conditions eq. (\ref{2eq4_5}). 
Consider the occurrence of equilibrium conditions. In order to ensure this,
the transitional semi-Markov dynamics should possess
an equilibrium distribution, i.e., there should exist the
densities $p_{\tau_1}^*(\tau_1)$, $p_{\tau_2}^*(\tau_2)$ satisfying the
stationary renewal equations
\begin{equation}
\frac{d p_{\tau_k}^*(\tau_k)}{d \tau_k}= - \nu_k(\tau_k) \, p_{\tau_k}^*(\tau_k) \, , \quad k=1,2
\label{2eq6_4}
\end{equation}
with
\begin{equation}
p_{\tau_k}^*(0)= \int_0^\infty \nu_k(\tau_k) \, p_{\tau_k}^*(\tau_k) \, d \tau_k
\label{2eq6_5}
\end{equation}
If $p_{\tau_k}^*(\tau_k)$, $k=1,2$ exist, then from eq. (\ref{2eq6_4}) it follows
that
\begin{equation}
p_{\tau_k}^*(\tau_k)= A \, \exp \left [ - \int_0^{\tau_k} \nu_k(\theta) \, d \theta \right ]
\label{2eq6_6}
\end{equation}
The latter condition implies that there exist constant $C>0$ and $\xi>1$, such that
for large $\tau_k$, say $\tau_k \gg 1$,
\begin{equation}
\exp \left [ - \int_0^{\tau_k} \nu_k(\theta) \, d \theta
 \right ] \leq \frac{C}{\tau_k^\xi} \,, \quad k=1,2
\label{2eq6_7}
\end{equation}
If these conditions are fulfilled, close to equilibrium, the density
$p(v,\theta,z,t;\tau_1,\tau_2)$ can be approximated by the factorized
form
\begin{equation}
p(v,\theta,z,t;\tau_1,\tau_2)= p_{\tau_1}^*(\tau_1) \, p_{\tau_2}^*(\tau_2) \,
p_{\bf v}(v,\theta,z,t)
\label{2eq6_8}
\end{equation}
Integrating eq. (\ref{2eq6_1}) with respect to $\tau_1$ and $\tau_2$,
one obtains for $p_{\bf v}(v,\theta,z,t)$ an equation identical
to eq. (\ref{2eq5_2}), where the constant rates $\nu_k$, $k=1,2$ entering
eq. (\ref{2eq5_2}) are substituted by the mean values  $\overline{\nu}_k$
\begin{equation}
\overline{\nu}_k= \int_0^\infty \nu_k(\tau_k) \, p_{\tau_k}^*(\tau_k) \,
d \tau_k \, , \quad k=1,2
\label{2eq6_9}
\end{equation}
Consequently, the values of the coefficients $q$, $b$ and $c$ fulfilling
the Langevin condition are
\begin{equation}
q= \sqrt{\frac{2 \, \gamma}{\overline{\nu}_1}} \,, \quad
c= - \sqrt{\frac{2}{\gamma \, \overline{\nu}_1}} \,, \quad  b= \sqrt{\frac{2 \, \lambda}{a \, \overline{\nu}_2}}
\label{2eq6_10}
\end{equation}
In a similar way, eqs. (\ref{2eq5_17}) generalize simply by substituting
$\overline{\nu}_\alpha^\prime$ and $\overline{\nu}_h$ to
$\nu_\alpha^\prime$ and $\nu_h$, $\alpha=1,\dots,N_i$, $h=1,\dots,N_d$.

\subsection{Examples}
\label{2sec5_1}
As an example of a semi-Markov impulsive realization of FD3k, consider the simple nondimensional
($\langle v^2 \rangle_{\rm eq}=1$)  dissipative
dynamics
\begin{equation}
\frac{d v}{d t}= - e^{-\lambda \, t} * v(t) + R(t)
\label{2ex2_1}
\end{equation}
where ``$*$'' indicates the operation of convolution, corresponding to eq. (\ref{2eq5_1})
in the presence of a single dissipative mode with relaxation rate $\lambda$, $a=1$, in the
absence of fluid inertial contributions, i.e. $\gamma=0$.

As a model for $\eta_2=\eta(t,\tau,\nu(\tau);\beta)$, as from eq. (\ref{2eq5_1}) we have in the
present case only a single dissipative modal stochastic forcing, consider a semi-Markov process 
with the transition rate given by eq.  (\ref{2eq4_14}), $\tau_0=1$, and the amplitude $\beta$ normally
distributed. Since  eq.  (\ref{2eq4_14}) is frequently encountered as a model for the superdiffusive
properties of L\'evy walks, the semi-Markov process $\eta(t,\tau,\nu(\tau);\beta)$ characterized by
(\ref{2eq4_14}) can be referred to as an impulsive semi-Markov process with a L\'evy transitional
structure.

\begin{figure}
\includegraphics[width=10cm]{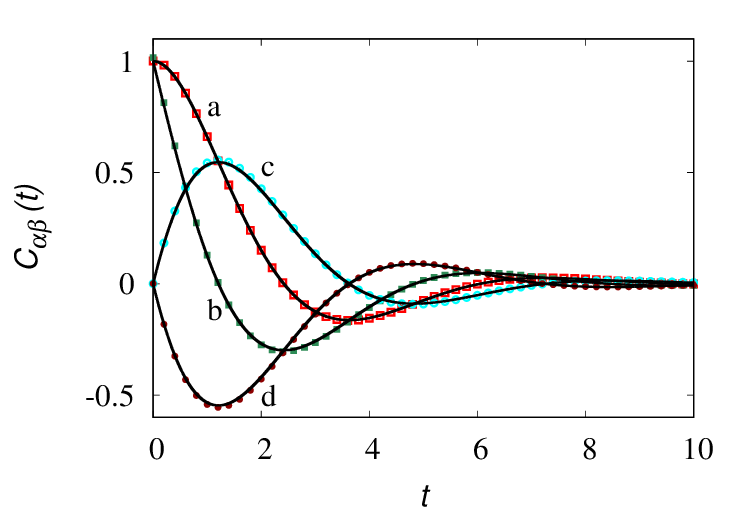}
\caption{Correlation functions  at equilibrium
for the model  eq. (\ref{2ex2_1})  with  $\lambda=1$,  driven by the
stochastic process $\eta(t,\tau,\nu(\tau),\beta)$, with L\'evy transition structure at $\xi=1.5$, 
and normal pdf $f_\beta(\beta)$
Symbols represent the results of stochastic simulations, solid lines correspond to the
result of LRT. Line (a): $C_{vv}(t)$, line (b): $C_{\theta \theta}(t)$, line (c): $C_{\theta v}(t)$,
line (d): $C_{v \theta}(t)$.}
\label{Fig_lw1}
\end{figure}

Figure  \ref{Fig_lw1} depicts the behaviour of the correlation
functions at equilibrium $C_{\alpha \beta}(t)$, where $\alpha, \, \beta=v, \,\theta$
obtained from stochastic simulations (using the same ensemble and averaging structure described for the Markovian
numerical simulations).

\begin{figure}
\includegraphics[width=10cm]{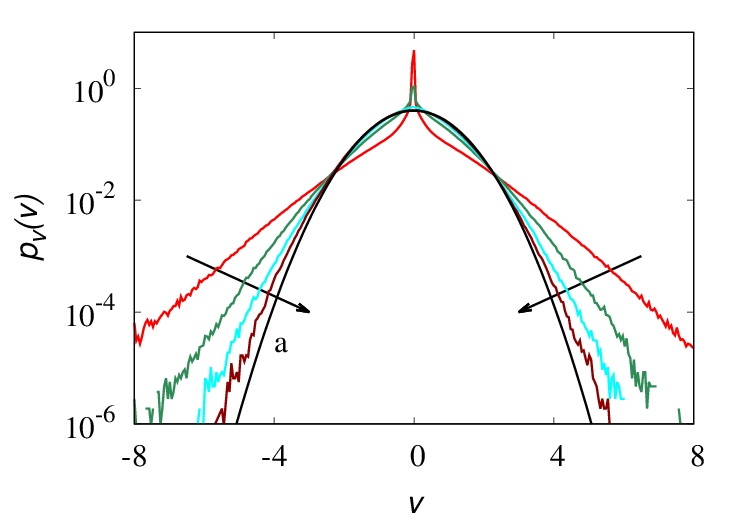}
\caption{Equilibrium velocity pdf $p_v(v)$ vs $v$ for   the model eq.  (\ref{2ex2_1})  
 with  $\lambda=1$,  driven by the
stochastic process $\eta(t,\tau,\nu(\tau),\beta)$, with L\'evy transition structure
and normal pdf $f_\beta(\beta)$ for several values of $\xi$. The arrow indicates increasing
values of $\xi=1.5,\,2.5,\, 4.5,\,10.5$. Line (a) represents the normal distribution.}
\label{Fig_lw2}
\end{figure}

The L\'evy transition model applies for $\xi>1$, and since $p_\tau(\tau)= (\xi-1)/(1+ \tau)^\xi$,
we have for the mean transition rate
\begin{equation}
\overline{\nu}= \langle \nu(\tau) \rangle_{p_\tau} = \xi \, (\xi-1)  \, \int_0^\infty \frac{d \tau}{(1+\tau)^{\xi+1}} =  \xi-1
\label{2ex2_2}
\end{equation}
Correspondingly,  the coefficient $b$ entering the modal representation of FD3k  implies
\begin{equation}
b= \sqrt{\frac{ 2 \, \lambda}{ \xi-1}}
\label{2ex2_3}
\end{equation}
The behaviour of the equilibrium velocity pdf is depicted in figure \ref{Fig_lw2},
taking $\xi$ as a parameter.

 The same qualitative phenomena observed in the Markovian case occur in semi-Markov models.
As $\xi$ increases, $\overline{\nu}$ increases, while $b$ decreases keeping fixed the ratio
$ \overline{\nu} \, b^2  \rightarrow 2 \, \lambda$, which is the same either in the
Markovian and in the semi-Markov cases, and for $\xi \rightarrow \infty$ the velocity pdf $p_v(v)$ progressively approaches  a Gaussian distribution.

\section{Generalized Fluctuational patterns}
\label{2sec6}

The analysis and the examples developed in the previous two Sections
can be integrated within the formulation of fluctuational patterns
developed in part I in a simple and  elegant way.

Consider again the impulsive Markov case, i.e. eqs. (\ref{2eq5_1}), (\ref{2eq5_6}),  (\ref{2eq5_15})
and its generalization eqs. (\ref{2eq5_16})-(\ref{2eq5_17}). The autocorrelation
function  of the impulsive process $\eta(t,\nu;\beta)$ i.e.  of the distributional
derivative of a compound Poisson process is given by
\begin{equation}
C_{\eta \eta}(t) = \langle \eta(t,\nu;\beta) \, \eta(0,\nu;\beta) \rangle =
\langle \beta^2 \rangle \, \nu \, \delta (t)
\label{2eqs_1}
\end{equation}
where
\begin{equation}
\int_{-\infty}^\infty C_{\eta \eta}(t) \, dt = \langle \beta^2 \rangle \, \nu 
\label{2eqs_2}
\end{equation}
Similarly for the semi-Markov counterpart $\eta(t,\tau, \nu(\tau);\beta)$  we have
\begin{equation}
C_{\eta \eta}(t) =  \langle \eta(t,\tau,\nu(\tau);\beta) \, \eta(0,\tau,\nu(\tau);\beta) \rangle   = 
\langle \beta^2 \rangle \, \overline{\nu} \, \delta (t)
\label{2eqs_3}
\end{equation}
where $\overline{\nu}$ is the mean value of $\nu(\tau)$.
Equations (\ref{2eq6_1}), (\ref{2eq6_3}) can be proved  by enforcing the Fourier approach developed in \cite{part3}
using Wiener-Khinchin theorem.

Let  $\widetilde{\eta}(t,\nu;\beta)$ and $\widetilde{\eta}(t,\tau,\nu(\tau);\beta)$ be the
normalized stochastic processes
\begin{equation}
\widetilde{\eta}(t,\nu;\beta)= \frac{\eta(t,\nu;\beta)}{\sqrt{\langle \beta^2 \rangle \, \nu}} \, , \qquad
\widetilde{\eta}(t,\tau, \nu(\tau);\beta)= \frac{\eta(t,\tau, \nu(\tau);\beta)}{\sqrt{\langle \beta^2 \rangle \, 
\overline{\nu}}}
\label{2eqs_4}
\end{equation}
possessing, by definition, unit value of the integral of the autocorrelation function,
\begin{equation}
\int_{-\infty}^\infty C_{\widetilde{\eta} \widetilde{\eta}}(t) \, d t = 1
\label{2eqs_5}
\end{equation}
If instead of $\eta(t,\nu;\beta)$ and $\eta(t,\tau,\nu(\tau);\beta)$,   we consider
the normalized processes $\widetilde{\eta}(t,\nu;\beta)$ or $\widetilde{\eta}(t,\tau, \nu(\tau);\beta)$,
it can be observed that  the dynamics  with impulsive  elementary forcings eqs. (\ref{2eq5_1}), (\ref{2eq5_16})
coincide  with the fluctuational
patterns introduced in part I   driven by Wiener processes,  
with $\xi_h(t)$ substituted by $\widetilde{\eta}_h(t,\nu;\beta)$ or
$\widetilde{\eta}_h(t,\tau, \nu(\tau);\beta)$, under the condition that the processes
  $\widetilde{\eta}_h$ are 
independent of each others. The patterns so constructed 
can be referred to as Generalized Fluctuational Patterns.
This result is formalized  by the following theorem, considering a generic fluctuational patterm with $N=N_d+N_i$ 
auxiliary internal
variables:\\

\noindent
{\bf Theorem I} - If the impulsive white-noise stochastic  processes $\widetilde{\eta}_h(t,\nu;\beta)$
or $\widetilde{\eta}_h(t,\tau, \nu(\tau);\beta)$, $h=1,\dots,N$ are mutually independent and 
characterized by the normalization condition eq. (\ref{2eqs_5}), then the generalized fluctuational
pattern obtained by substituting some or all of the  processes $\xi_h(t)$, $h=1,\dots,N$ with
$\widetilde{\eta}_h$ satisfies the Langevin condition and the fluctuation-dissipation relations
of the first and second kind.

\section{Hydromechanic theory of anomalous diffusion}
\label{2sec7}

The content of this Section, namely the hydromechanic theory of anomalous
diffusion is on the fringes of the analysis developed for non-Wiener stochastic forcings,
as we here consider the classical fluctuational patterns in the presence of Wiener stochastic processes.
The common denominator with the principal focus of the article is the analysis of
Gaussianity in  equilibrium velocity pdfs  in hydromechanic
situations that may give rise to anomalous diffusive motion.

The present  approach complements, on hydrodynamic bases, the scaling theory of anomalous
motion, that in the framework of the fluctuation-dissipation theory is referred to a GSER \cite{mason1,mason2,gser1}.

The starting point is the general linear hydromechanic   equation for the particle dynamics
defined by the dissipative kernel $h(t)$ and by  the fluid-inertial kernel $k(t)$,  thoroughly discussed
in part I, corresponding to the dynamic representation of the fluctuational
pattern ${\mathbb D} {\mathbb I}_{m,E}^{(1,N_d,N_i)}$, see part I. The only difference
with respect to part I is that either $N_d$ or $N_i$ (depending on the  phenomenology considered, i.e.
sub/superdiffusive motion) may be unbounded, so that the series representation of the kernel may
involve infinite terms.

To begin with consider  the scaling theory deriving from the particle equation of motion.
Let $\widehat{C}_{vv}(s)$ be the Laplace transform of the velocity autocorrelation function $C_{vv}(t)$.
We have from  FD1k, i.e. from the Langevin condition,
\begin{equation}
\widehat{C}_{vv}(s)= \frac{m \, \langle v^2 \rangle_{\rm eq}}{m \, s + \widehat{h}(s)+s \, \widehat{k}(s)}
\label{2eqan1}
\end{equation}
where $\widehat{h}(s)$ and $\widehat{k}(s)$ are the Laplace transforms of the dissipative and fluid-inertial
kernels, respectively.
Two cases may arise:
\begin{itemize}
\item either $\widehat{h}(0)=\int_0^\infty h(t) \, dt = \eta_\infty < \infty$, i.e. is  a finite number, and consequently
particle motion in the long-term is diffusively regular as it exists a diffusivity $D_\infty$ defined by the global
fluctuation-dissipation relation in conjunction with the Green-Kubo theorem,
\begin{equation}
D_\infty= \widehat{C}_{vv}(0)= \frac{m \, \langle v^2 \rangle_{\rm eq}}{\eta_\infty}
\label{2eqan2}
\end{equation}
\item or
\begin{equation}
\int_0^\infty h(t) \, d t = \infty
\label{2eqan3}
\end{equation}
and the diffusivity, defined according to eq. (\ref{2eqan2}), vanishes.
\end{itemize}
The second situation corresponds to subdiffusive anomalous diffusion and it is completely independent
of the nature of the fluid-inertial contribution, i.e. of $k(t)$. Eq. (\ref{2eqan3}) implies
that, for large $t$, $h(t) \sim t^{-\kappa}$, $\kappa \in (0,1)$.  This corresponds to
the scaling in the Laplace domain $\widehat{h}(s) \sim s^{\kappa-1}$ near $s=0$ and, since
$m \, s$ and $s \, \widehat{k}(s)$ are negligible with respect to  $\widehat{h}(s)$ for small $s$, we have
\begin{equation}
\widehat{C}_{vv}(s)=m \, \langle v^2 \rangle_{\rm eq} \, s^{1-\kappa} \, , \qquad s \rightarrow 0
\label{2eqan4}
\end{equation}
We can define a time-dependent diffusivity $D(t)=\int_0^t C_{vv}(\tau) \, d \tau$ (observe that $D_\infty =\lim_{t \rightarrow
\infty} D(t)$) so that $\widehat{D}(s)= \widehat{C}_{vv}(s)/s \sim s^{-\kappa}$.
As a consequence, the Laplace transform of the mean-square displacement  $\langle x^2(t) \rangle$ (assuming
that  initially  $x=0$) behaves near $s=0$ as
\begin{equation}
\int_0^\infty e^{-s t} \, \langle x^2(t) \rangle  \, d t= \frac{\widehat{D}(s)}{s} \sim \frac{1}{s^{1+\kappa}}
\label{2eqan5}
\end{equation}
implying for large $t$
\begin{equation}
\langle x^2(t) \rangle \sim t^{\kappa}
\label{2eqan6}
\end{equation}

\begin{table}
\begin{tabular}{c|c|c}
\hline
 & &  \\
$\;\;$ Dissipative kernel $h(t)$ $\;\;$ &  $\;\;$ Inertial kernel $k(t)$ $\;\;$ & $\;\;$ Long-term $\langle x^2(t) \rangle$  scaling$\;\;$\\
& &  \\
\hline
& &  \\
$\;$ $h(t) \neq 0,$ $\;\;$ $\displaystyle\int_0^\infty h(t) \, d t < \infty$ $\;\;$ & arbitrary & $\;\;$ Regular diffusion $\langle x^2(t) \rangle \sim t $ \\
&  &  \\
\hline
&  &  \\
$h(t) \sim t^{-\kappa},$  $\;\;$ $0 < \kappa < 1$ \;\; & arbitrary & $\;\;$ Subdiffusive regime $\langle x^2(t) \rangle \sim t^\kappa $ \\
& &  \\
\hline
  &  \\
$h(t)=0$ & $\;\;$ $k(t) \leq C \, t^{-\kappa},$  $\; C>0,\; \kappa> 1$ $\;\;$ & $\;\;$ Ballistic motion $\langle x^2(t) \rangle \sim t^2$ \\
& &  \\
\hline
&  &  \\
$h(t)=0$ & $\;\;$ $k(t) \sim t^{-\kappa},$  $\;\;$ $0 < \kappa < 1$ \;\; & $\;\;$ Superdiffusive regime $\langle x^2(t) \rangle \sim t^{1+\kappa} $ \\
& &  \\
\hline

\end{tabular}
\caption{Review of the scaling results  of the hydromechanic theory. The scaling relations for $h(t)$ and $k(t)$ refer to
the long-term limit $t \rightarrow \infty$.}
\label{tablescaling}
\end{table}

Therefore dissipative effect may determine  solely the transition from regular to anomalous subdiffusive  regimes.

There is a last situation that should be considered occurring in the
absence of dissipation, i.e. $h(t)=0$.
  Let us assume that, even in this particular case, equilibrium
conditions for the velocity dynamics are achieved.
This is indeed true for the  numerical  case studies addressed below, but the analysis of this
topic is far from being trivial and will be addressed elsewhere, as it is not strictly related to
the main goal of the present article.

Under these conditions  eq. (\ref{2eqan1}) reduces to
\begin{equation}
\widehat{C}_{vv}(s)= \frac{m \, \langle v^2 \rangle_{\rm eq}}{m \, s +s \, \widehat{k}(s)}
\label{2eqan7}
\end{equation}

Consider the case of a fluid-inertial kernel possessing
power-law tail for large $t$, i.e.
so that $\widehat{k}(s) \sim s^{\kappa-1}$, near $s=0$. Also in this case two situations occur
\begin{itemize}
\item either $\kappa >1$, or better to say more generally, there exists a positive constant $C>0$ and an exponent $\kappa > 1$,
such that $k(t) \leq C \, t^{-\kappa}$. In this case  the term $m s$ at the denominator of eq. (\ref{2eqan7}) prevails near $s=0$ with 
respect to $s \, k(s) \sim s^\kappa$. Thus $\widehat{C}_{vv}(s) \sim \langle v^2 \rangle_{\rm eq}/s$, and $\widehat{D}(s) \sim 1/s^2$,
the Laplace  transform of the mean square displacement scales as $1/s^3$, and this corresponds to
ballistic motion,
\begin{equation}
\langle x^2(t) \rangle \sim t^2
\label{2eqan8}
\end{equation}
\item or $\kappa <1$, the factor $s \, \widehat{k}(s)$ prevails on $m \, s$ near $s=0$, thus
\begin{equation}
\widehat{C}_{vv}(s) \sim \frac{m \, \langle v^2 \rangle_{\rm eq}}{s^\kappa}  \;\; \Rightarrow  \;\;
\widehat{D}(s) \sim \frac{1}{s^{1+\kappa}} \; \;  \Rightarrow  \;\;
\int_0^\infty e^{-s \,  t} \, \langle x^2(t) \rangle \, d t \sim \frac{1}{s^{2+\kappa}}
\label{2eqan9}
\end{equation}
that implies anomalous superdiffusive motion
\begin{equation}
\langle x^2(t) \rangle \sim t^{1+\kappa}
\label{2eqan10}
\end{equation}
\end{itemize}
Table \ref{tablescaling} reviews the scaling results obtained. 
The influence of the fluid-inertial kernel determines a transition from
 a ballistic motion,
corresponding to the absence of equilibrium conditions in the velocity 
dynamics, to a superdiffusive
motion. This occurs solely if  the relaxation of the fluid-inertial modes is sufficiently slow
to permit the achievement of equilibrium conditions. This phenomenon can be
easily explained by considering the  structure of the fluctuational pattern  ${\mathbb I}_{m,E}^{(1,1,N_i)}$
in the case the dissipation induced by the friction term is set equal to zero.
We can use the nondimensional formulation eq. (41) of part I (setting all the $a_h$'s equal to zero), and considering
$N_i=\infty$ (as otherwise it is not possible to achieve a long-term power-law scaling).
Under these assumptions, the dynamics  governed by frictionless fluid-inertial interactions
becomes the infinite system of equations
\begin{eqnarray}
\frac{d v}{d t} & =  & -  \left (\sum_{\alpha=1}^{\infty} \gamma_\alpha \right )  v 
+  \sum_{\alpha=1}^{\infty} \gamma_\alpha \, \mu_\alpha \, z_\alpha + \sqrt{2} \, \sum_{\alpha=1}^{\infty} d_\alpha \, \xi_\alpha^\prime(t) \nonumber \\
\frac{d z_\alpha}{d t}  &=  &- \mu_\alpha \, z_\alpha + v + \sqrt{2} \, c_\alpha \, \xi_\alpha^\prime(t) \,, \qquad \alpha=1,2,\dots,
\label{2eqan11}
\end{eqnarray}
For a slow inertial dynamics, the  apparent ``friction'' term $I_f(t)=-  \left (\sum_{\alpha=1}^{\infty} \gamma_\alpha \right )  v(t)$
(first term at the r.h.s. of eq. (\ref{2eqan11}))
permits the establishment of an equilibrium in the velocity dynamics, in the meaning that the opposite contribution, namely
$I_{cf}(t)=\sum_{\alpha=1}^\infty \gamma_\alpha \, \left [\mu_\alpha e^{-\mu_\alpha \, t} * v(t) \right ]$ (deriving
from the second term at the r.h.s. of eq. (\ref{2eqan11}) once the $z_\alpha(t)$-dynamics is substituted into it) is 
unable to equilibrate it in finite time. Of course at long time scales we have
\begin{equation}
\lim_{t \rightarrow \infty} \left [I_f(t) - I_{cf}(t) \right ] \, d t =0
\label{2eqan12}
\end{equation}
corresponding to the long-term absence of dissipation. From the physical point of view
the frictionless case may be of interest in the superfluidic case,   where the system
is characterized by a frictionless superfluidic phase of quantum nature \cite{superfl},
and more generally for problems at extremely low temperatures.

\begin{figure}
\includegraphics[width=10cm]{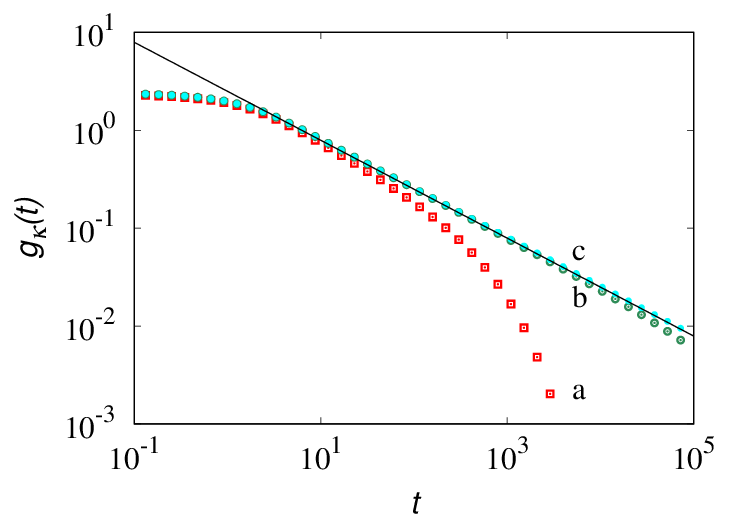}
\caption{Finite approximations of the function $g_\kappa(t)$  at $\kappa=0.5$ (solid line) for finite values of $N_i$:
(symbols $\square$, label a) $N_i=10$, (symbols $\circ$, label b) $N_i=20$, (symbols $\bullet$, label c) $N_i=40$.}
\label{Figan1}
\end{figure}

Let us consider some numerical simulations, by focusing attention on the frictlionless inertial case with $\kappa \in (0,1)$.
As a model for the inertial kernel we consider the function $g_\kappa(t)$ defined by the infinite
Prony series
\begin{equation}
g_\kappa(t) = \sum_{\alpha=1}^\infty \frac{1}{2^{\alpha \, \kappa}} \, e^{-t/2^\alpha}
\label{2eqan13}
\end{equation}
It is rather straightforward to check that in the long-term limit $g_\kappa(t) \sim t^{-\kappa}$.
Set $k(t)=g_\kappa(t)$, so that  with reference to eq. (\ref{2eqan11}) we have $\gamma_\alpha=2^{-\alpha \, \kappa}$
and $\mu_\alpha= 2^{-\alpha}$. In practical simulations the number $N_i$ of inertial modes should be truncated to
a finite number. Figure \ref{Figan1} depicts the approximation of $g_\kappa(t)$ by taking
a finite $N_i=10,\,20,\,40$ number of modes. It can be observed that the choice $N_i=20$ provides a good approximation
of the infinite sum up to nondimensional times order of $10^4$. For this reason
we choose this approximation as a convenient trade-off  between accuracy and computational time for numerically
solving the stochastic dynamics.

Simulations of particle hydromechanics have been obtained  by solving eq. (\ref{2eqan11}) with an Euler-Langevin algorithm
with a time step $\Delta t=10^{-3}$ using an ensemble of $10^6$ realizations 
 starting from
$v(0)=z_\alpha(0)=0$, $\alpha=1,\dots,N_i$ and $x(0)=0$ (in the estimate of the
mean square displacement). Once equilibrium conditions
have been achieved, the estimate of the equilibrium velocity density has been obtained from $10^8$ realizations, and the
estimate of the velocity autocorrelation function refers to the long-term conditions after a transient, once velocity statististics
has relaxed towards  equilibrium conditions.

\begin{figure}
\includegraphics[width=10cm]{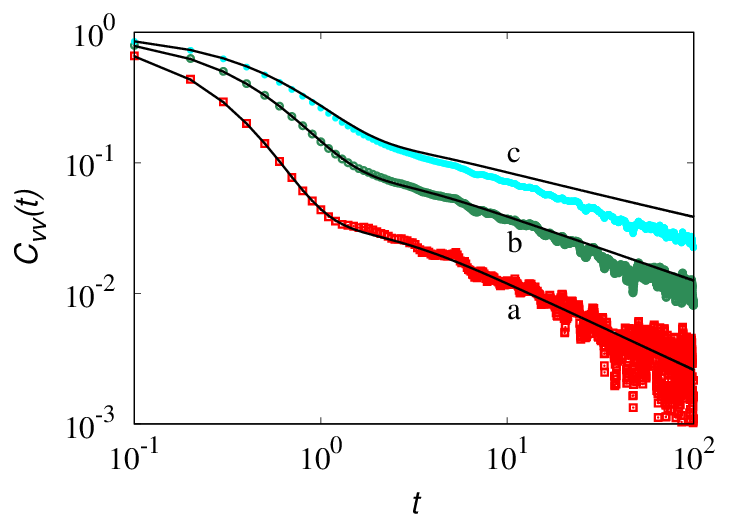}
\caption{$C_{vv}(t)$ vs $t$ for the frictionless inertial model discussed in the main text.
Solid lines represent the theoretical results deriving from FD1k, symbols the results of stochastic simulations:
(a) $\kappa=0.3$, (b) $\kappa=0.5$, (c) $\kappa=0.7$.}
\label{Figan2}
\end{figure}

\begin{figure}
\includegraphics[width=10cm]{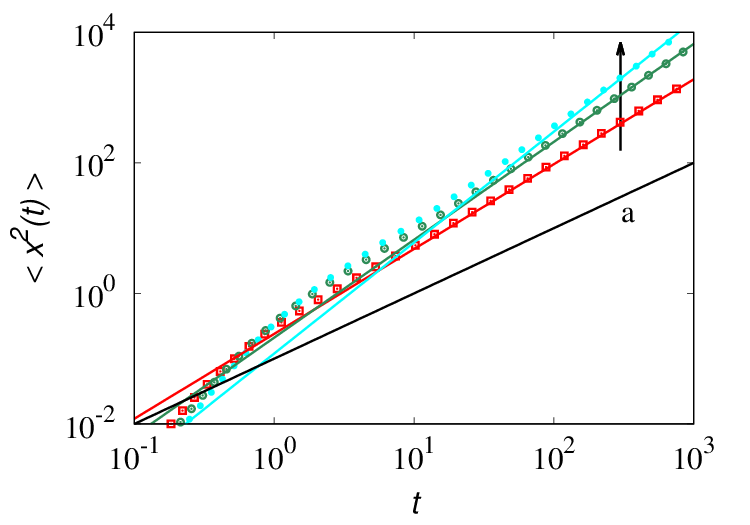}
\caption{$\langle x^2(t) \rangle$ vs $t$  for the frictionless inertial model discussed in the main text at different
values of $\kappa$. Symbols are the results of stochastic simulations. Line (a) represents the
regular scaling $\langle x^2(t) \rangle \sim t$ to mark the clear superdiffusive deviation from it.
The arrow indicates increasing values of $\kappa=0.3,\,0.5,\,0.7$, and the solid lines
the  scalings $\langle x^2(t) \rangle \sim t^{1+\kappa}$ at the corresponding value of $\kappa$.}
\label{Figan3}
\end{figure}

Figure \ref{Figan2} depicts the behaviour of the velocity autocorrelation function obtained from the
stochastic simulations (symbols) and from the theoretical FD1k result (solid lines) at different values of $\kappa$.
As expected $C_{vv}(t) \sim t^{-(1-\kappa)}$.
The behaviour of the mean-square displacements $\langle x^2(t) \rangle$ is depicted in figure \ref{Figan3} at different
values of $\kappa$. The supediffusive nature of motion is clearly evident.

\begin{figure}
\includegraphics[width=10cm]{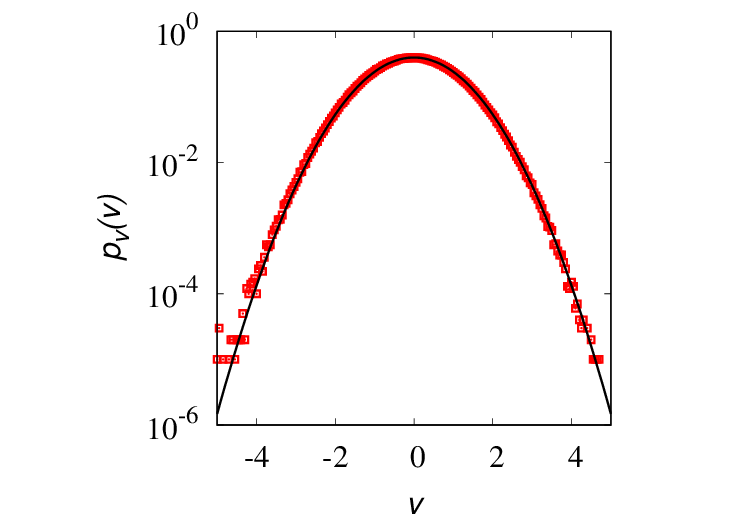}
\caption{Equilibrium velocity debsity function $p_v(v)$ vs $v$ for the frictionless inertial model discussed in the main text at 
$\kappa=.5$. Symbols are the results of stochastic simulations. The solid line is the normal
distribution $p_v(v)=e^{-v^2/2}/\sqrt{2 \pi}$.}
\label{Figan4}
\end{figure}

Next, consider the equilibrium velocity pdf. This is represented in figure \ref{Figan4} at $\kappa=0.5$
(other values of $\kappa$ provides the same result). Observe that within the present nondimensional
formulation $\langle v^2 \rangle_{\rm eq}=1$.
As can be observed, the collapsing towards a Gaussian distribution is excellent. And this
is generic features of all the hydromechanic models of anomalous diffusive 
behaviour, including
the dissipative subdiffusive case, not shown for the sake of conciseness.

Therefore, we may conclude that, despite diffusional anomalies,
 particle hydromechanics in the free-space
predicts in all the cases Gaussian velocity  density functions.

As a final remark, the frictionless inertial case represents a new  interesting model in the analysis of
nonequilibrium phenomena, as it show that it is possible to achieve equilibrium conditions (thermalization)
even in the absence of friction, providing that the inertial effects would relax sufficiently slow.
The thermodynamic implications of this results and a further analysis
 of this rather peculiar case will be addressed elsewhere, as it lies
outside the main scope of the present work.

\section{Concluding remarks}
\label{2sec8}
We have completed the analysis  initiated in
part I on the representation of FD3k relations within the classical Kubo paradigm
grounded on the validity of the Langevin condition.
The same block structure of the internal (dissipative, fluid-inertial) degrees
of freedom derived in part I for Wiener-driven modal forcings applies
 to generic models of stochastic forcings satisfying the condition of
fluctuational independence, which in practice corresponds to
the assumption of white modal stochastic perturbations.  This is clearly stated in
Theorem I in terms of fluctuational patterns.

The analysis developed in this article clearly shows that FD3k relations,
and specifically the representation of the thermal fluctuations is not unique,
if one considers exclusively the properties of the velocity autocorrelation function as the only relevant description of equilibrium fluctuations.
For a given $\langle v(t) \, v(0) \rangle_{\rm eq}$, there are uncountably many different
stochastic representations providing the same velocity autocorrelation 
function, making use   either of derivatives of Wiener processes,  or of
derivatives of Markovian and semi Markovian compound counting processes 
with arbitrary transition rates. 
This result suggests 
to extend the transport paradigm based exclusively on the
autocorrelation functions of the fluxes, by analyzing both theoretically
and experimentally higher-order correlation functions. This
observation is in accordance with the analysis developed in
\cite{gionaklagesfd} using a completely different formulation of
the Langevin equations.

Two other important issues stem from the analysis developed in the
present article. The use of distributional
derivatives of compount Markov and semi-Markov counting
processes unifies at the level of FD3k relations the classical Gaussian
analysis of thermal fluctuations with the definition of stochastic
perturbations arising in the study of athermal systems \cite{athermal1,athermal2}. In perspective, this suggests a further  conceptual
elaboration of the proper physical meaning of the concept of 
``athermality'', and this analysis could be extended to
the representation of stochastic fluctuations arising
in the statistical physical description of active systems and active
matter \cite{active}.

The second important point that follows from the results addressed
in this work is that Gaussianity should not be viewed as a law
of nature \cite{kacgauss} when dealing with thermal fluctuations.
It emerges as the most natural approximation in the presence of
frequent stochastic perturbations of small intensity, but it may
as well be violated in extreme physical conditions (e.g. at very low temperatures). This issue is closely connected with the analysis developed in
\cite{gionagaussian} where it is shown that Gaussianity, for dynamic variables,
such as particle/molecular velocities/momenta does not derive from a
mathematical principle (the Central Limit Theorem), but it is
a consequence of the elementary interaction mechanisms (amongst molecules,
or between molecules and radiation) in the most common physical conditions,
that may lead to a different thermodynamic behaviour if these conditions are suitably changed.

The hydromechanic theory of anomalous diffusion unifies in a single conceptual framework fluctuation-dissipation
phenomena possessing regular and anomalous diffusive properties.
Even in the anomalous case, the velocity fluctuations arising from Wiener-driven fluctuational patterns
possess a Gaussian pdf. It is also remarkable the extension of the theory to
the frictionless case and the onset of superdiffusive processes as a consequence of the
slow relaxational dynamics of fluid-inertial modes.\\

\vspace{0.2cm}

\noindent
{\bf Acknowledgment - } This research  received financial support from ICSC---Centro Nazionale di Ricerca in High Performance Computing, Big Data and Quantum Computing, funded by European Union---NextGenerationEU. One of the authors (M.G.) is grateful to
M. G.  Raizen for precious discussions.


\begin{thebibliography}{60}
\bibitem{part1}  M. Giona, G. Procopio and C. Pezzotti, Fluid-particle interactions and fluctuation-dissipation relations I -
General linear  theory and basic fluctuation patterns,  
submitted to arXiv (2024).

\bibitem{kubo} R. Kubo,  The fluctuation-dissipation theorem, Rep.  Prog.  Phys. {\bf 29}, 255-284 (1966).

\bibitem{kubolibro} R. Kubo, M. Toda and N. Hashitsune, {\em Statistical Physics II Nonequilibrium
Statistical Mechanics} (Springer-Verlag, Berlin, 1991).

\bibitem{procopvisco} G. Procopio and M. Giona, Modal representation of inertial effects in fluid-particle interactions and the regularity
of the memory kernels, Fluids {\bf 8}, 84 (2023).


\bibitem{ystatphys1} S. R. de Groot and P. Mazur, {\em Non-Equilibrium Thrmodynamics}
(Dover Publ., Mineola USA, 1984).

\bibitem{ystatphys2} L. Reichl, {\em A Modern Course in Statistical Physics} (Wiley-VCH,
Weinheim, 2009).
\bibitem{onsager1} L. Onsager and S. Machlup, Fluctuations and irreversible processes,
 Phys. Rev. {\bf 91}, 1505-1512 (1953).

\bibitem{onsager2}  L. Onsager and S. Machlup, Fluctuations and irreversible processes II. Systems with
kinetic energy, Phys. Rev. {\bf 91}, 1512-1515 (1953).

\bibitem{strato} R. L. Stratonovich, {\em Nonlinear Nonequilibrium Thermodynamics I:
Linear and Nonlinear Fluctuation-Dissipation Theorems} (Springer Verlag, Berlin,  1992).

\bibitem{athermal1} K. Kanazawa,  T. Sagawa and H.  Hayakawa,
Minimal Model of Stochastic Athermal Systems: Origin of Non-Gaussian Noise,
Phys. Rev. Lett. {\bf 114}, 090601 (2015).

\bibitem{athermal2} K. Kanazawa, T. G. Sano, T. Sagawa and H.  Hayakawa, 
Heat conduction induced by non-Gaussian athermal fluctuations, Phys. Rev. E 
{\bf 87}, 052124 (2013)

\bibitem{langevin}  P. Langevin,   Sur la theorie du mouvement brownien.
 C. R. Acad. Sci. (Paris)  {\bf 146}, 530-533 (1908), see also the English
translation in Am. J. Phys. {\bf 65}, 1079-1081 (1997).

\bibitem{chandra} S. Chandrasekhar,  Stochastic problems in physics and astronomy,
 Rev. Mod. Phys. {\bf 15}, 1-89  (1943).
\bibitem{vankampen} N. G. van Kampen, {\em Stochastic Processes in Physics and Chemistry}
(Elsevier, Amsterdam, 2007).
\bibitem{landauf} L. D. Landau and E. M. Lifshitz {\em Fluid Mechanics} (Pergamon Press, Oxford UK, 
1993).
\bibitem{gionaklagesfd} M. Giona, D. Cocco, G. Procopio, A. Cairoli and R. Klages,
Stochastic hydrodynamic velocity field and the representation of Langevin equations,  arXiv:2302.11672 (2023).

\bibitem{Franosch} F. Höfling and T. Franosch, Anomalous transport in the crowded world of biological cells, 
Rep.  Prog.  Physics {\bf 76}, 046602 (2013).

\bibitem{Fedotov1} S. Fedotov and N. Korabel, Subdiffusion in an external potential: Anomalous effects hiding behind normal behavior, 
Phys. Rev. E {\bf 91}, 042112 (2015).

\bibitem{Regner}  B. M. Regner, D. Vučinić, C. Domnisoru, T. M.  Bartol, M. W. Hetzer, D. M. Tartakovsky and T. J. Sejnowski, 
Anomalous diffusion of single particles in cytoplasm, Biophys. J.  {\bf 104}, 1652-1660 (2013).




\bibitem{mason1} T. G. Mason, Estimating the viscoelastic moduli of complex fluids using the generalized Stokes–Einstein equation, 
Rheol. Acta {\bf 39}, 371-378 (2000).

\bibitem{mason2} T. G. Mason Mason, K. Ganesan, J. H. van Zanten, D. Wirtz and S. C.  Kuo,
 Particle tracking microrheology of complex fluids, Phys. Rev. Lett. {\bf 79}, 3282-3285 (1997).



\bibitem{mason3}  T. M. Squires and T. G. Mason, Fluid mechanics of microrheology, Ann. Rev.Fluid Mech. {\bf 42}, 413-438 (2010).


\bibitem{gser1} W. Hong, G.  Xu, X. Ou, W. Sun, T. Wang and Z. Tong, 
Colloidal probe dynamics in gelatin solution
during the sol–gel transition, Soft Matter {\bf 14}, 3694-3703 (2018).
\bibitem{gser2} K. Suman and Y. M. Joshi, On the universality of the scaling relations during sol-gel transition,
J. Rheol. {\bf 64}   863-877 (2020).
\bibitem{gser3} Y. Dai, R. Zhang, W. Sun, T. Wang, Y. Chen and Z.  Tong, Z,  Dynamical heterogeneity in the gelation process of a polymer solution with a lower critical solution temperature. Soft Matter {\bf 17}, 3222-3233 (2021).

\bibitem{McKinley2018} S. A. McKinley and H. D. Nguyen, Anomalous diffusion and the generalized Langevin equation, SIAM Journal 
 Math. Anal {\bf 50}, 5119-5160 (2018).

\bibitem{Ferreira2022} R. M. S. Ferreira, From generalized Langevin stochastic dynamics to anomalous diffusion, Phys. Rev. E 
{\bf 106}, 054157 (2022).

\bibitem{Tuladhar2017} R. Tuladhar, M. Bologna, and P. Grigolini, Non-Poisson renewal events and memory, Phys. Rev. E {\bf 96}, 042112 (2017).


\bibitem{semimarkovcount1} N. Laskin, Fractional Poisson process, Comm. Nonlin. Sci.
 Num. Sim. {\bf 8}, 201-213 (2003).

\bibitem{semimarkovcount2} D. Cocco  and M. Giona, 
Generalized counting processes in a stochastic environment, Mathematics 
{\bf 9},  2573 (2021).

\bibitem{gardiner} C. Gardiner, {\em Stochastic Methods} (Springer-Verlag, Berlin, 2009).
\bibitem{white2} C. van Den Broeck, On the Relation between White Shot Noise, Gaussian
White Noise, and the Dichotomic Markov Process, J. Stat. Phys. {\bf 31}, 467-483 (1983).
\bibitem{bedeaux1} P. Mazur and D. Bedeaux,  Causality, Time-Recersla Invariance and the
Langevin Equation, Physica A {\bf 173}, 155-174 (1991).
\bibitem{bedeaux2} P. Mazur and D. Bedeuax, Nature of the Random Force in Brownian Motion,
Langmuir {\bf 8}, 2947-2951 (1992).
\bibitem{hanggirel} J. Dunkel and  P. H\"anggi, 
Relativistic brownian motion, Phys. Rep. {\bf 471},  1-73 (2009).
\bibitem{compoundpoisson} W. Feller, {\em In Introduction to Probability Theory and its Applications
- Volume 1} (J. Wiley \& Sons, New York, 1968).
\bibitem{fedotov} S. Fedotov, Single integrodifferential wave equation for a L\'evy walk,
 Phys. Rev.  E {\bf 93},  020101 (2016).
\bibitem{jphysA} M. Giona, M. D'Ovidio, D. Cocco, A. Cairoli and
R. Klages, Age representation of Lévy walks: partial density waves, 
relaxation and first passage time statistics,
 J.  Phys. A {\bf 52},  384001 (2019).
\bibitem{extendedPK} M. Giona,  A. Cairoli and R. Klages,
"Extended Poisson-Kac theory: A unifying framework for stochastic processes with finite 
propagation velocity, Phys. Rev. X {\bf 12},  021004 (2022).
\bibitem{levy1} V. Zaburdaev, S. Denisov and Joseph Klafter,
L\'evy walks, Rev.  Mod. Phys.  {\bf 87}, 483-530 (2015).
\bibitem{kac} M. Kac, A stochastic model related to the telegrapher's equation,
 Rocky Mountain J.  Math. {\bf 4}, 497-509 (1974).
\bibitem{gpk1} M. Giona,  A. Brasiello, and S. Crescitelli,
 Stochastic foundations of undulatory transport phenomena: Generalized 
Poisson–Kac processes—Part I basic theory,
 J.  Phys. A {\bf 50}, 335002 (2017).
\bibitem{part3}  M. Giona, G. Procopio and C. Pezzotti,
Fluid-particle interactions and fluctuation-dissipation relations III -
Correlated fluctuations, regularity and added mass, submitted to arXiv (2024).
\bibitem{superfl} A: Schitt, Introduction To Superfluidity (Springer, Cham (CH), 2015).
\bibitem{active} P. Romanczuk, M. B\"ar, W. Ebeling, B. Lindner and L. Schimansky-Geier,
Active Brownian particles: From individual to collective stochastic dynamics,
Eur. Phys. J. ST {\bf 202}, 1-162 (2012).
\bibitem{kacgauss} M. Kac {\em Statistical Independence in Probability, Analysis  \&
Number Theory} (Dover Publ., Mineola USA, 2018).
\bibitem{gionagaussian} M. Giona, C. Pezzotti  and  G. Procopio,
 Another normality is possible. Distributive transformations and emergent Gaussianity,
Physica A {\bf 634}, 129450 (2024).
\end{thebibliography}
\end{document}